\documentclass[manuscript]{elsarticle}
\usepackage{booktabs}
\usepackage[xindy,style=index,numberedsection=false]{glossaries}
\usepackage{siunitx}
\usepackage{subcaption}
\usepackage{paralist}
\usepackage[inline]{enumitem}
\usepackage{setspace}
\usepackage[hidelinks]{hyperref}
\usepackage{cleveref}
\usepackage{amsfonts}
\usepackage{xcolor}
\usepackage{soul}
\soulregister\gls7
\soulregister\glspl7
\soulregister\cite7
\soulregister\Cref7
\soulregister\cref7

\usepackage[ruled]{algorithm2e}

\usepackage{algorithmicx}

\crefname{algocf}{algorithm}{algorithms}
\Crefname{algocf}{Algorithm}{Algorithms}

\journal{Journal Of Parallel And Distributed Computing}

\begin{document}
\newacronym{sae}{SAE}{Sparse Approximate Eigenproblem}
\newacronym{evd}{EVD}{Eigenvalue Decomposition}
\newacronym{ff}{FF}{Flip Flop}
\newacronym{lut}{LUT}{Look Up Tables}
\newacronym{sa}{SA}{Systolic Array}
\newacronym{nlp}{NLP}{Natural Language Processing}
\newacronym{ppr}{PPR}{Personalized PageRank}
\newacronym{spmv}{SpMV}{Sparse Matrix-Vector Multiplication}
\newacronym{topkspmv}{Top-K SpMV}{Top-K Sparse matrix-vector multiplication}
\newacronym{coo}{COO}{Coordinate}
\newacronym{dsl}{DSL}{Domain-Specific Language}
\newacronym{slr}{SLR}{Super Logic Region}
\newacronym{csc}{CSC}{Compressed Sparse Column}
\newacronym{csr}{CSR}{Compressed Sparse Row}
\newacronym{raw}{RAW}{Read-After-Write}
\newacronym{vram}{VRAM}{Video RAM}
\newacronym{dram}{DRAM}{Dynamic RAM}
\newacronym{ndcg}{NDCG}{Normalized Discounted Cumulative Gain}
\newacronym{imp}{IMP}{Indirect Memory Prefetcher}
\newacronym{ir}{IR}{Information Retrieval}
\newacronym{dcg}{DCG}{Discounted Cumulative Gain}
\newacronym{er}{ER}{Entity Resolution}
\newacronym{rob}{ROB}{Reorder Buffer}
\newacronym{ii}{II}{Initiation Interval}
\newacronym{gpu}{GPU}{Graphics Processing Unit}
\newacronym{fpga}{FPGA}{Field Programmable Gate Array}
\newacronym{colamd}{COLAMD}{ColumnApproximate Minimum Degree algorithm}
\newacronym{cuda}{CUDA}{Compute Unified Device Architecture}
\newacronym{cpu}{CPU}{Central Processing Unit}
\newacronym{hbm}{HBM}{High Bandwidth Memory}
\newacronym{ddr}{DDR}{Double Data Rate}
\newacronym{bscsr}{BS-CSR}{Block-Streaming CSR}
\newacronym{glove}{GloVe}{Global Vectors for Word Representation}
\newacronym{cu}{CU}{Compute Unit}
\newacronym{pe}{PE}{Processing Element}
\newacronym{sp}{SP}{Stream Processor}
\newacronym{uram}{URAM}{UltraRAM}
\newacronym{cordic}{CORDIC}{Coordinate Rotation Digital Computer}
\newacronym{iram}{IRAM}{Implicitly Restarted Arnoldi Method}
\newacronym{dsa}{DSA}{Domain Specific Architecture}
\newacronym{llc}{LLC}{Last Level Cache}
\newacronym{ldm}{LDM}{Latent Diffusion Model}
\newacronym{nn}{NN}{Neural Network}
\newacronym{lm}{LM}{Language Model}
\newacronym{ml}{ML}{Machine Learning}
\newacronym{sve}{SVE}{Scalable Vector Extension}
\newacronym{numa}{NUMA}{Non-Uniform Memory Access}
\newacronym{ig}{IG}{Information Gain}
\newacronym{cmg}{CMG}{Core Memory Group}
\newacronym{armpl}{ARMPL}{Arm Performance Libraries}
\newacronym{dcsr}{DCSR}{Doubly Compressed Sparse Row}
\newacronym{ell}{ELL}{ELLPACK}
\newacronym{pmu}{PMU}{Performance Monitor Unit}
\newacronym{cdf}{CDF}{Cumulative Distribution Function}
\newacronym{spadd}{SpADD}{Sparse Matrix Addition}
\newacronym{spgemm}{SpGEMM}{Sparse General Matrix-Matrix Multiplication}
\newacronym{gcp}{GCP}{Google Cloud Platform}
\newacronym{aws}{AWS}{Amazon Web Services}
\newacronym{oci}{OCI}{Oracle Cloud Infrastructure}
\newacronym{hpc}{HPC}{High Performance Computing}
\newacronym{mshr}{MSHR}{Miss Status Holding Register}
\newacronym{mape}{MAPE}{Mean Absolute Percentage Error}
\newacronym{pmc}{PMC}{Performance Monitoring Counter}
\newacronym{mpki}{MPKI}{Misses Per Kilo Instruction}
\newacronym{alu}{ALU}{Arithmetic Logic Unit}
\newacronym{pim}{PIM}{Processing In Memory}
\newacronym{gflops}{GFLOPS}{Giga Floating Point Operations Per Second}
\newacronym{mlp}{MLP}{Memory Level Parallelism}
\newacronym{noc}{NOC}{Network On Chip}
\newcommand{\graviton}{Graviton 3}
\newcommand{\kp}{Kunpeng 920}
\newcommand{\postk}{A64FX}
\newcommand{\spmv}{\gls{spmv}}
\newcommand{\spadd}{\gls{spadd}}
\newcommand{\spgemm}{\gls{spgemm}}
\newcommand{\arm}{Arm}
\newcommand{\XXX}{\textcolor{red}{XXX}}

\newcommand{\marco}[1]{\textcolor{black}{#1}}
\newcommand{\ivan}[1]{\textcolor{black}{#1}}
\newcommand{\miquel}[1]{\textcolor{black}{#1}}
\newcommand{\adria}[1]{\textcolor{black}{#1}}
\newcommand{\fra}[1]{\textcolor{black}{#1}}

\newcommand\ivancmt[1]{\noindent{\color{green} {\bf \fbox{ivan: }} {\it#1}}}
\newcommand\fracmt[1]{\noindent{\color{blue} {\bf \fbox{fra: }} {\it#1}}}

\newcommand{\forreview}[1]{\textcolor{blue}{#1}}


\title{SpChar: Characterizing the Sparse Puzzle via Decision Trees}


\author[bsc,upc]{Francesco Sgherzi\corref{ca}} 
\ead{francesco.sgherzi@bsc.es}

\author[bsc,upc]{Marco Siracusa}
\ead{marco.siracusa@bsc.es}

\author[bsc,upc]{Ivan Fernandez}
\ead{ivan.fernandez@bsc.es}
\author[bsc,upc]{Adrià Armejach}
\ead{adria.armejach@bsc.es}
\author[bsc,upc]{Miquel Moretó}
\ead{miquel.moreto@bsc.es}
\affiliation[bsc]{%
            organization={Barcelona Supercomputing Center},
            addressline={Plaça d'Eusebi Güell, 1-3}, 
            city={Barcelona},
            postcode={08034}, 
            country={Spain}}

\affiliation[upc]{%
            organization={Universitat Politècnica de Catalunya},
            addressline={Carrer de Jordi Girona, 31}, 
            city={Barcelona},
            postcode={08034}, 
            country={Spain}}

\cortext[ca]{Corresponding author}
            
\glsunset{cpu}
\begin{abstract}
Sparse matrix computation is crucial in various modern applications, including large-scale graph analytics, deep learning, and recommender systems. The performance of sparse kernels varies greatly depending on the structure of the input matrix, making it difficult to gain a comprehensive understanding of sparse computation and its relationship to inputs, algorithms, and target machine architecture. Despite extensive research on certain sparse kernels, such as \spmv{}, the overall family of sparse algorithms has yet to be investigated as a whole. This paper introduces SpChar, a workload characterization methodology for general sparse computation.
SpChar employs tree-based models to identify the most relevant hardware and input characteristics, starting from hardware and input-related metrics gathered from \glspl{pmc} and  matrices. Our analysis enables the creation of a \textit{characterization loop} that facilitates the optimization of sparse computation by mapping the impact of architectural features to inputs and algorithmic choices. We apply SpChar to more than 600 matrices from the SuiteSparse Matrix collection and three state-of-the-art \arm{} \glspl{cpu} to determine the critical hardware and software characteristics that affect sparse computation. In our analysis, we determine that the biggest limiting factors for high-performance sparse computation are (1) the latency of the memory system, (2) the pipeline flush overhead resulting from branch misprediction, and (3) the poor reuse of cached elements.
Additionally, we propose software and hardware optimizations that designers can implement to create a platform suitable for sparse computation. We then investigate these optimizations using the gem5 simulator to achieve a significant speedup of up to 2.63$\times$ compared to a \gls{cpu} where the optimizations are not applied.
\end{abstract}

\label{sec:abstract}

\maketitle

\glsresetall
\section{Introduction}\label{sec:intro}
Operating with sparse matrices is crucial for many of today's applications. Kernels like \spmv, \spgemm, and \spadd{} are the building blocks for Recommender Systems~\cite{parravicini2021scaling}, Ranking~\cite{parravicini2021reduced}, Genomics~\cite{rakocevic2019fast} and, more broadly, serve as a proxy for the operations performed in the more broader Sparse Tensor Algebra field~\cite{baskaran2012efficient}. 
Moreover, as ever bigger Language Models ~\cite{brown2020language} and Latent Diffusion Models~\cite{rombach2022high} enter mass adoption, it has become apparent that both training and deploying \textit{dense} models incur in extremely high machinery and energy costs~\cite{sharir2020cost, strubell2019energy}, as well as hindering the ability to deploy such models on-edge rather than in a data center. In this scenario, Neural Network sparsification has been proven effective in reducing energy and time requirements for training and inference~\cite{zhou2021effective,peng2022towards,zhou2021efficient,hoefler2021sparsity}. 

Sparse workloads exhibit diverse memory access patterns and arithmetic intensities, which vary
greatly across kernels as well as inputs, thereby making input characterization crucial to
determine the requirements for a platform to excel in sparse computation.
On a fundamental level, Sparse kernels greatly benefit from high bandwidth memories in
conjunction with low latency memories, as most of them exhibit a scan-and-lookup access
pattern~\cite{umuroglu2016random}. However, memory systems typically fail to deliver the stringent bandwidth and latency demands of sparse workloads. For this reason, the available compute throughput is hindered by the memory system~\cite{giannoula2022sparsep}, thus requiring complex prefetching mechanisms~\cite{byna2008taxonomy}, and large caches to better utilize the core's arithmetic capabilities. 

Orthogonally, determining the impact of a certainal architectural feature (e.g., size of \gls{mshr}, memory technology, number of cores) on an kernel applied to a given input is crucial in sectors that employ matrices of  specific kinds, which is the case in the fields of structural engineering, social network mining and sparse neural networks.
In this setting, a promising approach that is gaining traction is the use of \gls{ml} models~\cite{malakar2018benchmarking,wu2022survey} to estimate the performance and impact of an architectural change quickly, in contrast to more accurate but slower simulation-based approaches~\cite{lowe2020gem5,wang2022evaluation}.

While performance characterization of some sparse kernels, namely \spmv, has been thoroughly explored ~\cite{nisa2018effective,benatia2016machine,chen2020characterizing,alappat2022execution}, the problem of choosing an appropriate architecture for sparse computation (hereafter referred as \textit{sparse problem}) has yet to be tackled as a whole, even if other sparse kernels like \spadd{} or \spgemm{} are fundamental operations for the broader tensor algebra field. Moreover, recent work that characterizes those kernels mostly focuses on a single architecture and a single kernel, while only employing limited sets of matrices to carry the evaluation forward.
As a result, architectural insights on sparse computation are hard to obtain due to the interdependence of architecture, inputs, and kernels in this category of heavily data-dependent problems.

SpChar is the first to tackle the \textit{sparse problem} from a holistic point of view.
Specifically, our goal is to create a new analysis method to establish a \textit{characterization loop} that enables hardware and software designers to map the impact of architectural features to algorithmic choices and input, and then optimize from the insights gathered. 
To this end, we gather matrices from 9 different domains, and we statically characterize (i.e., without running any sparse kernels) them based on the impact they have on certain architectural features (e.g., ease of branch prediction, randomness of the resulting access pattern, imbalance when partitioned on multiple threads). We use them to benchmark \spmv, \spgemm, and \spadd{} on 3 different \arm{} platforms and profile \adria{the executions using} \glspl{pmc} \adria{via} the \texttt{perf} suite.
We then feed both the matrix metrics and a subset of the \glspl{pmc} to a Decision Tree Regressor and extract the most relevant splitting attributes. By comparing splitting attributes across \glspl{cpu} we can see the impact of architectural choices and inputs on a given kernel.
In our analysis, we show that (1) \spmv{} is bottlenecked by the latency of the memory system when inputs show low locality and by the overhead of pipeline flushes \adria{due to} branch mispredictions, (2) \spgemm{} is bottlenecked by continuous cache evictions as a result of poor reuse of the values of the right-hand side matrix, and (3) \spadd{} is bottlenecked by the branch misprediction overhead and is less dependent \adria{on} the underlying matrix structure.

In summary, this paper presents the following contributions: 
\begin{itemize}
    \item We present SpChar: a workload characterization methodology for Sparse computation that considers both the hardware perspective and the input perspective. We employ tree-based models determine what are the most impactful \gls{cpu} features (e.g., size of the \gls{llc}, memory technology) for a given a combination of kernel and input.
    \item We evaluate our methodology on a set of 600 matrices spanning 9 different domains (e.g., structural engineering, social networks) using 3 different ARMv8 \glspl{cpu} on 3 sparse kernels.
    \item We explore a practical application of SpChar within a software-hardware characterization loop. To this end, we assess the effectiveness of our proposed optimizations on the \spmv{} kernel, executed on a simulated \arm{} platform. Our proposed optimizations are effective in reducing the execution time and increasing throughput of \spmv{}, with a speedup of up to 2.63$\times$ compared to a \gls{cpu} where such optimizations are not applied.
\end{itemize}

\section{Background and Motivation} \label{sec:motivation_and_background}
\subsection{Sparse Formats and Kernels} \label{sec:background}
\begin{figure}
    \centering
    \includegraphics[width=0.8\linewidth]{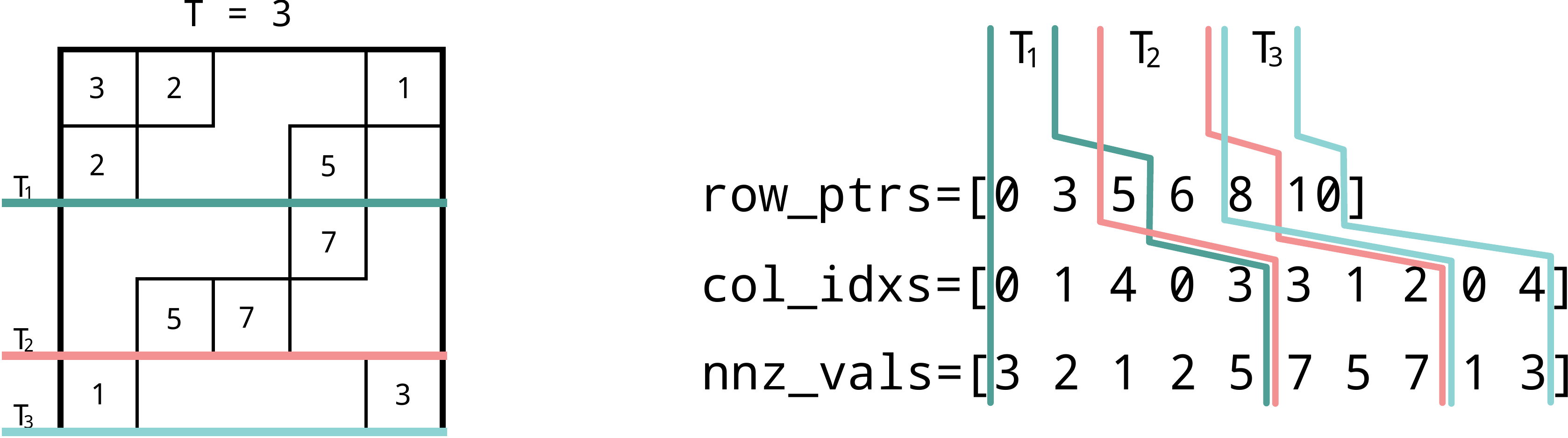}
    \caption{Row-wise partitioning scheme for \gls{csr} on 3 threads. Each thread operates on the contiguous set of rows displayed on the left, which translates to the \textit{influence regions} on the right. \fra{Note that partitions of the \texttt{row\_ptrs} array overlap because each thread requires knowing what is the index of the last element the previous thread operated on. }}
    \label{fig:matrix_partitioning}
\end{figure}
This section introduces the sparse formats and kernels we selected to study sparse computation. As sparse linear-algebra libraries are less mature compared to their dense counterparts, we could only select three kernels with a single format. Moreover, we wanted to select simple-enough kernels that just contain few operations in order to study their performance impact in isolation.

\subsubsection{The CSR format}
Although there are many sparse formats in the literature, \gls{csr}~\cite{barret94csr,im01block_csr} is perhaps one of the most used~\cite{langr16survey_sparse_formats}. As depicted in \Cref{fig:matrix_partitioning}, it is composed of three arrays:
\begin{itemize}
    \item \texttt{nnz\_vals}: stores the values of the non-zero entries of a matrix in row-major order.
    \item \texttt{col\_idxs}: stores the column indexes of the non-zero entries of a matrix in row-major order.
    \item \texttt{row\_ptrs}: stores the position of the first element of each row within the \texttt{nnz\_vals} and \texttt{col\_idxs} arrays.
\end{itemize}
Overall, CSR provides a good trade-off between storage efficiency and versatility. In fact, the \texttt{row\_ptrs} vector provides direct row indexing, which is essential in certain dataflow algorithms (e.g., SpGEMM CSR \cite{gustavson78spgemm}) and row-based work partitioning as shown in \Cref{fig:matrix_partitioning}. 

\subsubsection{The SpMV kernel}\label{sec:background_spmv}
As detailed in \Cref{alg:spmv}, \gls{spmv}~\cite{williams07spmv} multiplies a sparse matrix with a dense vector and returns a dense vector. 
If the sparse matrix stores graph adjacencies, this nested-loop structure is a typical traversal operation that can be found in many graph algorithms that \gls{spmv} is the building block of (e.g. Pagerank, Breadth First Search, etc) \cite{gapbs15beamers}.

When using \gls{csr}, thread parallelism is generally enforced row-wise on the outer loop whereas data parallelism is enforced element-wise in the inner loop~\cite{gomez2021efficiently}.
On today's processors, the most critical operation of this algorithm is the indirect access on the dense vector~\cite{yang21spzip,im01block_csr}.
If the access pattern has low spatio-temporal locality, elements are likely to be gathered from lower cache layers, requiring longer-latency accesses that saturate the memory subsystem.
Another important aspect is that the dependent nested-loop structure performing the matrix traversal introduces a row overhead that may harm performance on matrices with few non-zeros per row~\cite{kanellopoulos2019smash}.
As indirect access and traversal of compressed sparse formats are common operations in sparse linear algebra and tensor-algebra kernels (e.g. SpMM \cite{yang2018design}, MTTKRP \cite{kurt2022sparsity}, etc), we believe SpMV is one of the most important proxies to study sparse computation. 

\subsubsection{The SpGEMM kernel}\label{sec:background_spgemm}
\noindent
\begin{minipage}[t]{.46\textwidth}
    \begin{algorithm}[H]
        \setstretch{0.85}
        \caption{\gls{spmv} \gls{csr} algorithm}\label{alg:spmv}
        \KwData{$A \in \mathbb{R}^{pxq} \textsf{, } x \in \mathbb{R}^{q}$}
        \KwResult{$y \in \mathbb{R}^{p}$}
        \Comment{$A$ is a sparse matrix, $y, x$ are dense vectors}
        
        \For{$a_{i*} \in A$} {
            \For{$a_{ij} \in a_{i*}$} {
                $y_{i} \gets y_{i} + a_{ij} \times x_{j}$\;
            }
        }
    \end{algorithm}
\end{minipage}%
\hspace{.3cm}
\begin{minipage}[t]{.51\textwidth}
    \begin{algorithm}[H]
        \setstretch{0.85}
        \caption{SpGEMM \gls{csr} algorithm}\label{alg:spgemm}
        \KwData{$A \in \mathbb{R}^{pxq} \textsf{, } B \in \mathbb{R}^{qxr}$}
        \KwResult{$C \in \mathbb{R}^{pxr}$}
        \Comment{$A, B, C$ are sparse matrices}
        
        \For{$a_{i*} \in A$} {
            \For{$a_{ij} \in a_{i*}$} {
                \For{$b_{jk} \in b_{j*}$} {
                    $value \gets a_{ij} \times b_{jk}$\;
                    \If{$c_{ik} \notin c_{i*}$}{
                        $c_{ik} \gets 0$\;
                    }
                    $c_{ik} \gets c_{ik} + value$\;
                }
            }
        }
\end{algorithm}
\end{minipage}

\gls{spgemm} multiplies two sparse matrices and returns a sparse matrix.
This algorithm is of fundamental importance for linear algebra and graph analytics as is widely used for multigrid solvers, triangle counting, multi-source BFS, and others~\cite{gao20spgemm_survey}.

Although there are several versions of \gls{spgemm}, we only consider the Gustavson's implementation~\cite{gustavson78spgemm}
reported in \Cref{alg:spgemm}, as the mostly used one~\cite{gao20spgemm_survey}.
Conversely from \gls{spmv}, \gls{spgemm} also requires to build the output matrix that is generally generated in two phases: symbolic (firstly populates \texttt{row\_ptrs} and allocates \texttt{col\_idxs} and \texttt{nnz\_vals}) and numeric (then computes and writes \texttt{col\_idxs} and \texttt{nnz\_vals}).
In both cases, this algorithm sequentially traverses matrix $A$ and indirectly accesses rows of matrix $B$ , which are then accumulated and stored.
We believe \gls{spgemm} is an interesting test case as (1) the indirect access on matrix $B$ has more spatial locality but less temporal locality than \gls{spmv}, stressing the memory subsystem in a different way and (2) the accumulation operation is of fundamental importance in sparse computation and is also used in higher-dimensional kernels such as sparse tensor contraction.

\subsubsection{The SpADD kernel} \label{sec:background_spadd}

\gls{spadd} adds two sparse matrices and returns a sparse matrix.
As reported in \Cref{alg:spadd}, the kernel iterates through the rows of the two matrices and merges them (copy if the given index is present in only one row, sum otherwise) if they both contain some non-zero element or just copies them otherwise \cite{hussain2021spkadd}.
In contrast to the other kernels, \gls{spadd} is not as memory intensive since the two CSR matrices are traversed sequentially.
However, the disjunctive-merging operation is quite control intensive, particularly stressing \glspl{alu} and branch predictors.
Since merging operations --- both in disjunctive and conjunctive mode --- are quite common in higher-order tensor algebra (e.g. sparse tensor contraction) \cite{fred2017taco}, we believe \gls{spadd} is a representative proxy to study the control overhead of sparse operations.

\subsection{The case for Workload Characterization of Sparse Algorithms} \label{sec:workload_char}
In recent years, the application of sparse algorithms has evolved from being limited to scientific codes~\cite{ng2001spectral,coleman1983estimation,el2007hardware} and benchmarks \cite{dongarra2015hpcg,stathis2003d} to being ubiquitous in today's workloads. With the rise of ever-bigger graphs representing social network topologies and more detailed user-buy-product characterizations for recommender systems, it is clear that sparse computation is gradually expanding to the cloud domain. Cloud providers like \gls{gcp}, \gls{aws}, Alibaba Cloud, and \gls{oci} already provide different \glspl{cpu} to best suit the needs of their clients. With all of them now offering \arm{} alternatives to the more mainstream X86\_64 machines and even building their own (i.e., Graviton series for \gls{aws} and Yitian 710 for Alibaba Cloud), the application domain of \arm{} architectures is expanding from mobile to \gls{hpc} and datacenters. 
\begin{algorithm}[t]
\setstretch{0.85}
\caption{\gls{spadd} \gls{csr} algorithm}\label{alg:spadd}
\KwData{$A \in \mathbb{R}^{pxq} \textsf{, } B \in \mathbb{R}^{pxq}$}
\KwResult{$C \in \mathbb{R}^{pxq}$}
\For{$a_{i*} \in A \textsf{ and } b_{i*} \in B$} {
  \hskip\algorithmicindent \textbf{merge (copy or sum)} $a_{i*}$ \textbf{and} $b_{i*}$ \textbf{until some row is fully scanned}  \par
  \hskip\algorithmicindent \textbf{copy remaining elements in} $a_{i*}$ \par
  \hskip\algorithmicindent \textbf{copy remaining elements in} $b_{i*}$
}
\end{algorithm}

Now more than ever, algorithms operating on sparse matrices are employed in a plethora of domains, many of which share very few similarities. This discrepancy stems from sparse algorithms being vastly dependent on the underlying matrix structure in contrast to algorithms operating on dense matrices being generally input agnostic.
Moreover, as we discuss in \Cref{sec:background}, the degree to which the performance of an algorithm depends on the input dramatically depends on the algorithm itself: algorithms like \spmv{} and \spgemm{} are inherently more sensitive than \spadd.

In light of this, while architectural features to optimize dense arithmetic operations are well studied and characterized for \glspl{cpu}~\cite{andreev2015vectorization,carneiro2021lightweight}, \glspl{gpu}~\cite{navarro2020gpu,haidar2018harnessing}, and \glspl{dsa}~\cite{de2020fblas,kara2017fpga}; it is not immediately clear which are the design decision that would best benefit sparse computation. Traditionally, having higher \gls{mlp} has made \glspl{gpu} the \textit{de facto standard} platform for sparse computation; however, higher memory latency~\cite{martineau2019benchmarking} and smaller caches make them less applicable to workloads that operate on limited size matrices, like recommender systems~\cite{parravicini2021scaling} or graph clustering~\cite{sgherzi2022eigen}, with a possible solution being \gls{cpu}-\gls{gpu} co-computation~\cite{benatia2020sparse}.
In the scope of performance characterization of sparse workloads, while there exists an extensive body of work characterizing \spmv{} ~\cite{elafrou2018sparsex,giannoula2022sparsep,nisa2018effective,benatia2016machine,chen2020characterizing,alappat2022execution,siracusa2020roofline}
and the structure of the inputs~\cite{elafrou2017performance,goumas2008understanding,im2004sparsity}, \spadd{} and \spgemm{} have historically received less attention, despite being relevant proxies for the operations performed in the broader tensor algebra field~\cite{chen2020aesptv,tian2021high,chou2022dynamic}. To this day, advancements in the algorithms above are mainly carried forward from the algorithmic side~\cite{buluc2008challenges,borvstnik2014sparse, niu2022tilespgemm}, with limited attention to the inputs' structure~\cite{ballard2016hypergraph}.

To the best of our knowledge, there exists no prior work characterizing several sparse matrix algorithms on different architectures that focuses deeply on the inputs' structure.

\subsection{Mapping Architectural Features to Inputs and Algorithms}\label{sec:motivation_mapping}
While \gls{ml} methods are starting to gain traction in design space exploration and performance prediction, they are still not well established in the broader field of computer architecture. This is due to them having limited explainability (see Burkart et al.~\cite{burkart2021survey} and references therein), thus making the task of automatizing the gathering of architectural insights from the model parameters far from trivial. Exceptions to this rule are, among others~\cite{sinha2013multivariate,buckley2014generalised,hastie87gam,kokel2020unified}, tree-based models~\cite{loh2011classification}, which have seen adoption in hardware-software codesign ~\cite{singh2019napel,ould2007using,calder1997evidence}.

In addition to being \textit{more explainable}, tree-based models are paving the way for the use of \gls{ml} in computer architecture due to being generally more resilient to the magnitude of the input features and fairly easy to use and deploy. Recent works have used tree-based models for performance prediction~\cite{tousi2022comparative,bodin2016integrating} and automatic design space exploration~\cite{hutter2011sequential,cianfriglia2018model}. Within the scope of extracting relevant software/architectural insight from profiling, Fenacci et al.~\cite{fenacci2010workload} employ decision trees to gather insights on benchmarks targeting embedded applications. Albeit focusing on the performance characterization of specific domains (e.g. networking, telecommunications, automotive), they employ only data derived by performance counters and do not take into account input-related metrics. As a result, while domains are characterized from the hardware perspective, seeing how workloads that are data-dependent react to the variation of inputs and how this interacts with hardware features is not achievable with this method. 
More recently Bang et al. \cite{bang2020hpc} has used tree-based models to determine relevant features of IO-intensive workloads starting from logs generated from the top 20 apps executed on the CORI Supercomputer. As a result, the insights they gather are more focused on the general behavior of the suite of applications deployed on the supercomputer rather than on hardware characteristics.

Moreover, within the scope of extracting relevant hardware/software features from decision trees, one needs to exercise an abundance of caution when drawing the conclusion that some splitting attribute of the decision tree impacts the architecture in a relevant way, as these methods reflect the correlations present in the dataset rather than implying causality between features. As we later discuss in \Cref{sec:relevant_insights_extraction}, one way to escape the correlation-implies-causation dilemma is to compare the relevant attributes from different models targeting the same features, analyze the presence (and absence) of features across models and ultimately link them to the known architectural choices for a given machine.

To the best of our knowledge, no prior work exists that uses decision trees to extract relevant hardware and input insights by comparing different architectures.

\section{SpChar Methodology}\label{sec:methodology}
\glsunset{ddr}
\glsunset{hbm}
\glsunset{ell}
\glsunset{coo}
\glsunset{dcsr}
\subsection{Hardware and Algorithms} \label{sec:hardware_and_algos}
We profile and benchmark three state-of-the-art \arm{} platforms from both Cloud and HPC sectors: Fujitsu's \postk{} \cite{sato2020co}, Huawei's \kp{} \cite{xia2021kunpeng}, and the more recent \graviton{} \cite{graviton3}
used on AWS. Such platforms are not only a good representation of the \arm{} ecosystem at the time of writing this paper, but they also have different memory technologies, as well as greatly different architectural design choices in terms of cache sizes, size of the vector units, and core count. \Cref{tab:cpus} summarizes the most important features of the architectures under study and their system software.
\glsunset{cmg}
\begin{table}[!ht] 
    \centering
    \caption{Summary of the architectural and software features of the \glspl{cpu}}

  \resizebox{\textwidth}{!}{%
    \begin{tabular}{@{}llll@{}}
    \toprule
       &\textbf{\postk} & \textbf{\kp} & \textbf{\graviton}  \\ \midrule
        Manufacturer & Fujitsu & Huawei & Amazon \\ 
        Architecture & ARM v8.2 & ARM v8.2 & ARM v8.4 \\ 
        Sockets & 1 & 2 & 1 \\ 
        Cores per socket & 48 & 64 & 64 \\
        Vector units & 2 $\times{}$ 512-bit SVE & 1 $\times{}$ 128-bit NEON & 2 $\times{}$ 256-bit SVE \\
        L1D, L1I per core & \SI{64}{\kilo\byte}, \SI{64}{\kilo\byte} & \SI{64}{\kilo\byte}, \SI{64}{\kilo\byte}  & \SI{64}{\kilo\byte}, \SI{64}{\kilo\byte} \\ 
        Private L2 &  N/A & 64 $\times{}$ \SI{512}{\kilo\byte} & 64 $\times{}$ \SI{1}{\mega\byte}  \\ 
        Shared LLC & 4 $\times{}$ \SI{8}{\mega\byte} & \SI{64}{\mega\byte} & \SI{32}{\mega\byte} \\ 
        Memory technology & \gls{hbm}2 & \gls{ddr}4 & \gls{ddr}5 \\ 
        Memory channels & 32 & 16 & 8 \\ 
        Peak bandwidth & \SI{1024}{\giga\byte / \second} & \SI{380}{\giga\byte / \second} & \SI{300}{\giga\byte / \second} \\ \midrule
        Operating System & Red Hat Enterprise Linux 8.1 & CentOS Linux 7 & Ubuntu 22.04 \\
        Compiler & \texttt{armclang} 22.0.2 & \texttt{armclang} 22.0.2 & \texttt{armclang} 22.0.2 \\ 
        Kokkos Version & \texttt{Kokkos 3.5} & \texttt{Kokkos 3.5} & \texttt{Kokkos 3.5} \\ 
        \bottomrule
    \end{tabular}}
    \label{tab:cpus}
\end{table}

We benchmark \spmv{}, \spadd{} (Symbolic/Numeric), and \spgemm{} (Symbolic/Numeric) kernels using single precision floating point for the \texttt{nnz\_vals} vector and unsigned integers for the \texttt{row\_ptrs} and the \texttt{col\_idxs} vectors (\SI{4}{\byte} wide in our systems). The kernels are provided by the Kokkos Kernels~\cite{rajamanickam2021kokkos} library, which we choose as it provides implementations that are vectorized both for NEON~\cite{reddy2008neon} (\kp) and \gls{sve}~\cite{stephens2017arm} (\postk{} and \graviton). \fra{For all input matrices, we run the specific kernel in two phases. First, we run the kernel $J$ times without profiling it, which removes the effect of spinning up \texttt{OpenMP} threads and warms up hardware structures (e.g., caches). After the warmup phase, we run the kernel $K$ times, profiling each run individually. In our experiments, we use $J = 10$ and $K = 5$, which strikes a balance between having profiling results with low variance across runs and having reasonable execution times. While Kokkos Kernels provides an implementation of \spmv{} that is \textit{ready to use}, some options need to be tweaked for \spadd{} and \spgemm{}. For starters, both algorithms need the \texttt{team\_work\_size} to be set, which refers to the number of fundamental operations (row sum for \spadd{}, row product for \spgemm{}) that each thread is assigned per OpenMP batch.
For our benchmarks, we set this value to $16$, as it is the same value used by the authors of the library for performance regression testing in version $3.5$. While higher values (e.g., $32$, $128$, $256$) might yield better performance in matrices distributed with low imbalance across threads, it also decreases performance significantly in matrices with high imbalance. Choosing the optimal value requires \textit{a priori} knowledge of the input matrix, and different architectures might perform optimally under distinct \texttt{team\_work\_size} parameters, even for the same matrix. As our goal is evaluating different architectures, we deem more relevant to choose sensible defaults rather than optimizing ad-hoc for each combination of matrix and architecture. Regarding
\spgemm{}, the library proposes several algorithmic choices. For our evaluation, we select the \texttt{KK\_SPEED} version as it yields better performance at the cost of an higher memory usage.}

While high-performance implementations of such algorithms are also present in the \gls{armpl}, the implementation of \spadd{} is not multithreaded and both \spadd{} and \spgemm{} do not expose methods to execute the symbolic phase by itself. We release the code for this benchmarking suite and can be found at \href{https://gitlab.bsc.es/fsgherzi/spchar-benchmarking}{https://gitlab.bsc.es/fsgherzi/spchar-benchmarking}. 

We use the \gls{csr} matrix format~\cite{greathouse2014efficient}, as it is widely employed and supported by several libraries \cite{lane2023heterogeneous,eo2022roofline,flegar2017overcoming,kim2022analysis}, including Kokkos. While a plethora of other sparse matrix formats do exist (e.g. \gls{coo}, \gls{dcsr}, \gls{ell} and variants), they are either (1) domain or architecture specific~\cite{parravicini2021scaling,maggioni2013adell}, (2) could incur in wasteful data replication~\cite{chen2022msrep}, or (3) require additional decompression at computation time~\cite{willcock2006accelerating}. \gls{csr}, on the other hand, is fairly agnostic of the matrix structure/domain, does not require decompressing data before being used, and opens itself to simple vectorization techniques. 

\subsection{Profiling Methodology}\label{sec:profiling}
We select \texttt{perf} \cite{de2010new} as the profiling framework as it comes pre-packaged with most Linux distributions and is relatively easy to use. Operating with performance counters can be performed via plain syscalls (\texttt{perf\_event\_open}), which has streamlined the instrumentation of the benchmarks. Other profiling frameworks were considered (\texttt{likwid} \cite{treibig2010likwid}, \texttt{papi} \cite{dongarra2001using}, \texttt{extrae} \cite{eriksson2016profiling}) but later discarded due to either offering no additional functionality compared to \texttt{perf}, being too fine-grained (trace-based profiling), offering aggregated metrics, or being more complex in usage within the scope of the instrumentation of the kernels. Nonetheless, profiling results were cross-referenced frequently to determine the correctness of the ad-hoc \texttt{perf} instrumentation.

Performance counter identifiers for  architectures can be obtained from two distinct sources: the official \arm{} \gls{pmu} reference and the \gls{cpu} manufacturer optimization guides. 
The \arm{} \gls{pmu} references contain counters that are core-related (e.g., frontend and backend stalls, core cache misses, scalar and vector floating point operations), whereas the counters provided by the manufacturers generally cover \adria{uncore} architectural elements, such as the network-on-chip, last level caches, memory controllers and communication to memories and data exchanges across \gls{numa} domains.
In our analysis, we use the counters provided by the manufacturer reference optimization guides for \postk{} and \kp{}, whereas we use the \arm{} \gls{pmu} reference for \graviton{} as no document containing uncore counters is publicly available at the time of writing this document.
The list of performance counters for each \gls{cpu} can be found in \Cref{tab:pmu}.

\begin{table}[!ht] 
    \centering
    \caption{Summary of \gls{pmu} counters for each \gls{cpu}. \textbf{Counter ID} refers to the name used in the \gls{pmu} reference manuals of \postk, \kp, \graviton.}
    \begin{tabular}{@{}lll@{}}
    \toprule
       & \textbf{Counter ID} & \textbf{Supported \gls{cpu}} \\ \midrule
        Cycles elapsed &  \texttt{CPU\_CYCLES} &  All  \\
        Executed Instructions & \texttt{INST\_RETIRED} &  All \\
        Cycles stalled in frontend & \texttt{STALL\_FRONTEND} &  All \\
        Cycles stalled in backend & \texttt{STALL\_BACKEND} &  All \\
        Scalar fp instructions & \texttt{VFP\_SPEC} &  All \\
        Vector fp instructions & \texttt{ASE\_SPEC} &  All \\
        Load cacheline in L2 & \texttt{L2\_CACHE\_REFILL} &  \postk \\
        Store cacheline to memory & \texttt{L2\_CACHE\_WB} &  \postk \\
        LLC Miss (L2) & \texttt{L2\_MISS\_COUNT} & \postk \\
        LLC Miss (L3) & \texttt{L3\_CACHE\_MISS\_RD} & \kp, \graviton \\
        L1 Accesses   & \texttt{L1D\_CACHE} & All \\ 
        Memory Accesses & \texttt{MEM\_ACCESS} & \kp, \graviton \\
        Branch mispredicted & \texttt{BR\_MIS\_PRED} & All \\
        Branch executed & \texttt{BR\_RETIRED} & All \\ 
        \bottomrule
    \end{tabular}
    \label{tab:pmu}
\end{table}

\subsection{Datasets}\label{sec:dataset}
The kernels are tested on the biggest 600 matrices by number of nonzero elements from the SuiteSparse Matrix collection \cite{suitesparse}. These matrices have varying sizes ($\sim$ [$10^6$, $10^9$] nonzeros) and densities ($\sim$ [$10^{-7} \%, 25\%$]), and come from 9 different domains (e.g., computer vision, network problems, social networks), thus exhibiting different structures that are representative of real-world fields of applications. \fra{The biggest 600 matrices were chosen in order to prevent our \glspl{cpu} to fully cache computational elements (e.g., right-hand side matrix in \spadd{} and \spgemm{}, dense vector in \spmv{}) at any one time. }

In addition to real-world matrices, we enrich our dataset with synthetic matrices, specifically generated to stress a particular \gls{cpu} characteristic. Albeit the number of nonzero elements of each matrix varies depending on the generation method, the number of rows and columns is fixed at 16 million to impede the presence of computational elements like the dense vector (\SI{64}{\mega \byte}) for \spmv{} in the last-level cache.

\begin{figure}
    \centering
       \begin{subfigure}{0.25\textwidth}
        \centering
        \includegraphics[width=\textwidth]{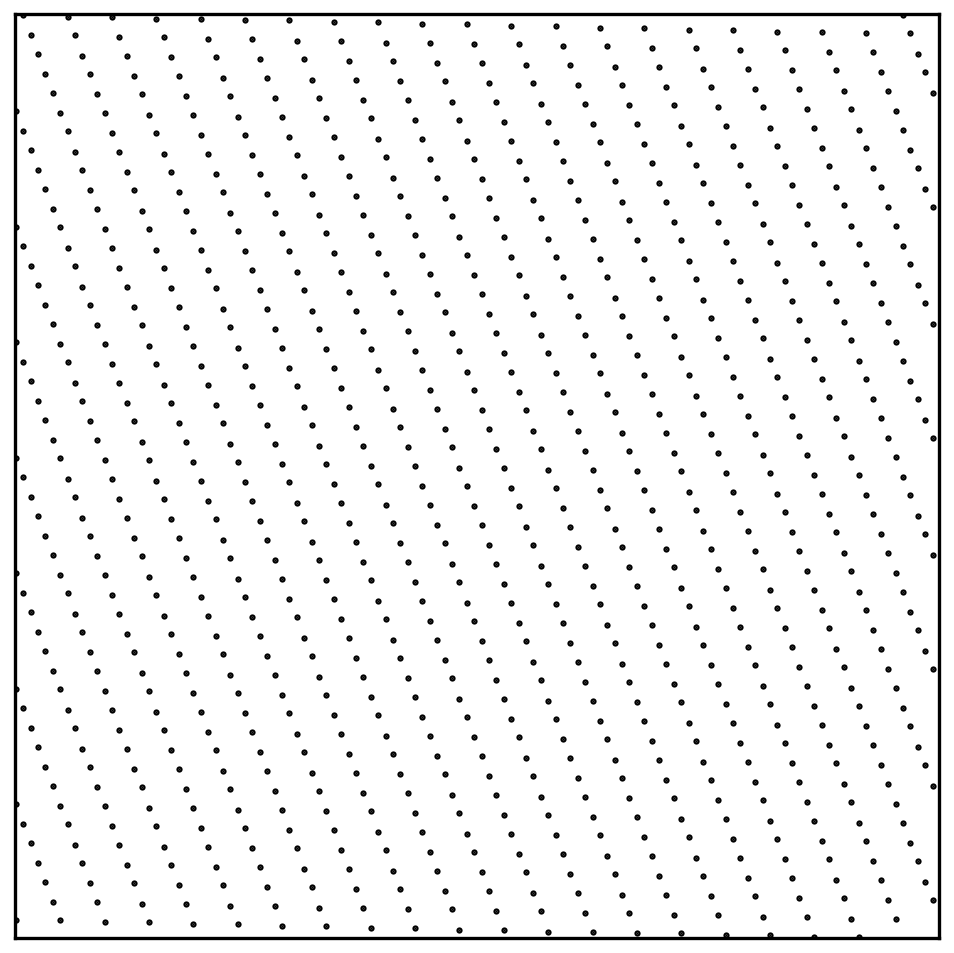}
        \caption{Stride}
        \label{fig:mat_stride}
    \end{subfigure}%
    \begin{subfigure}{0.25\textwidth}
        \centering
        \includegraphics[width=\textwidth]{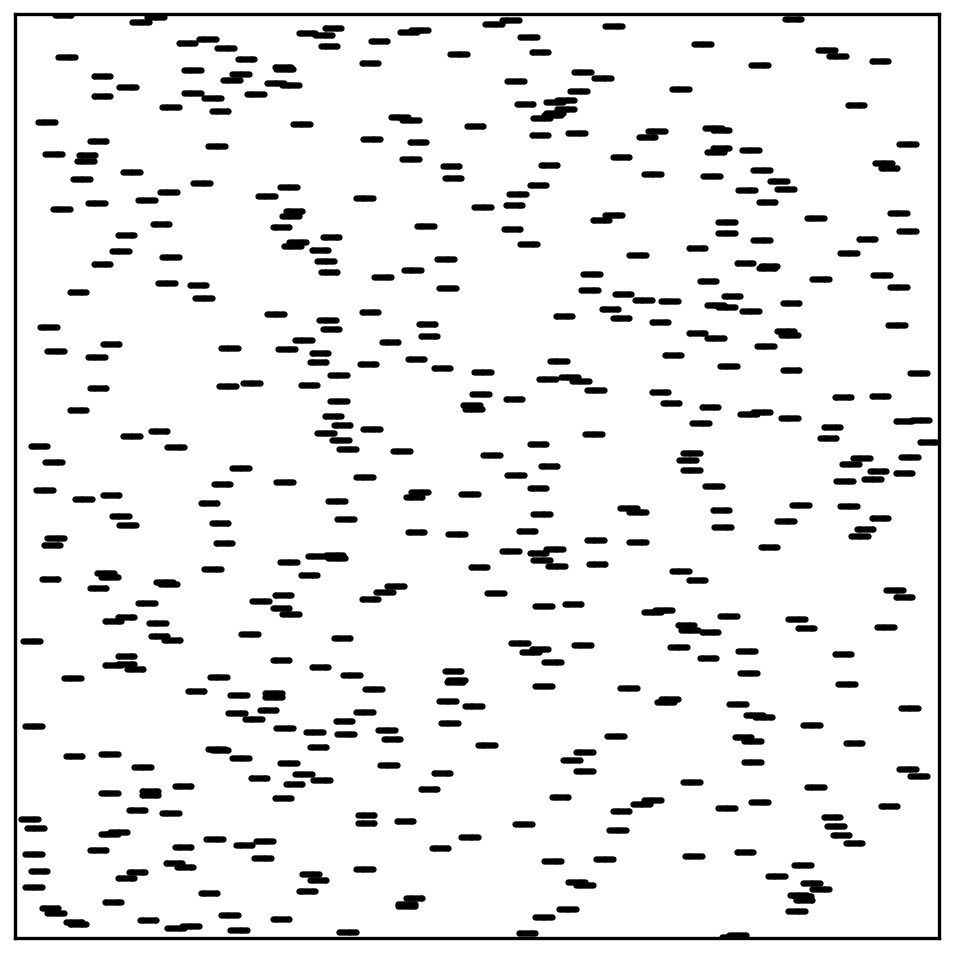}
        \caption{Spatial loc.}
        \label{fig:mat_spatial}
    \end{subfigure}%
    \begin{subfigure}{0.25\textwidth}
        \centering
        \includegraphics[width=\textwidth]{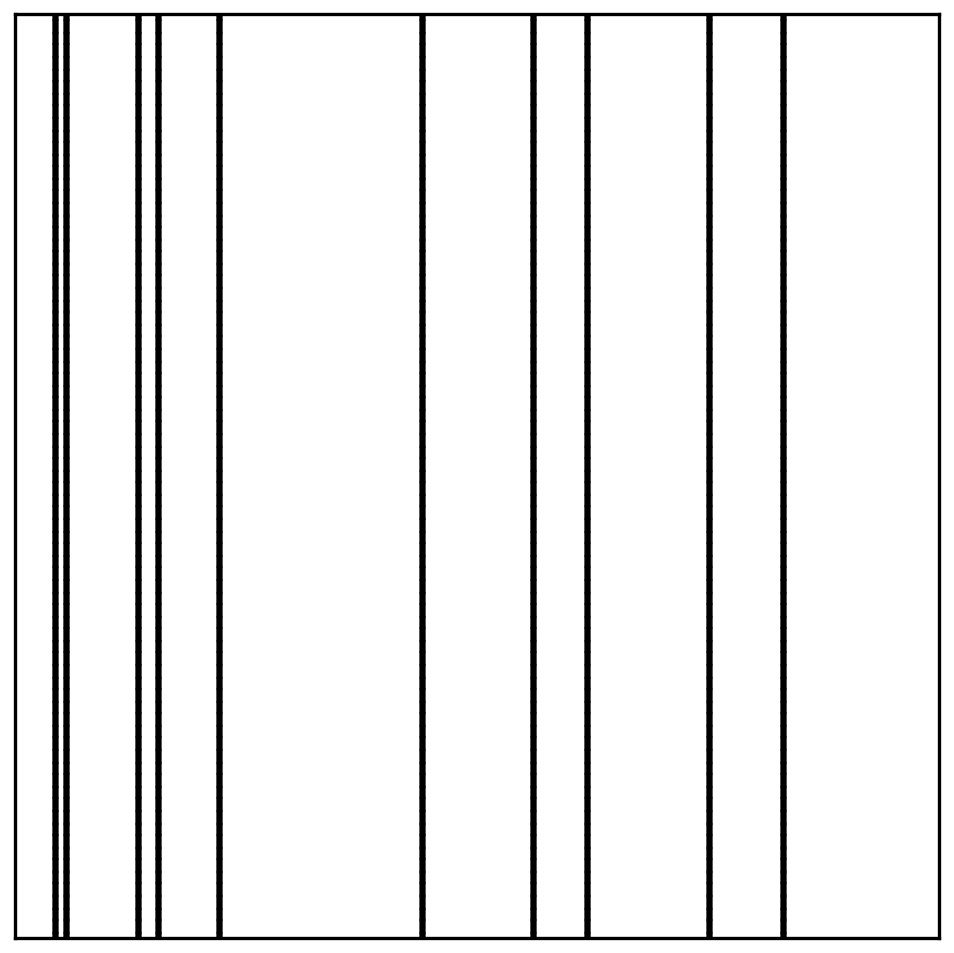}
        \caption{Temporal loc.}
        \label{fig:mat_temporal}
    \end{subfigure}%
    \begin{subfigure}{0.25\textwidth}
        \centering
        \includegraphics[width=\textwidth]{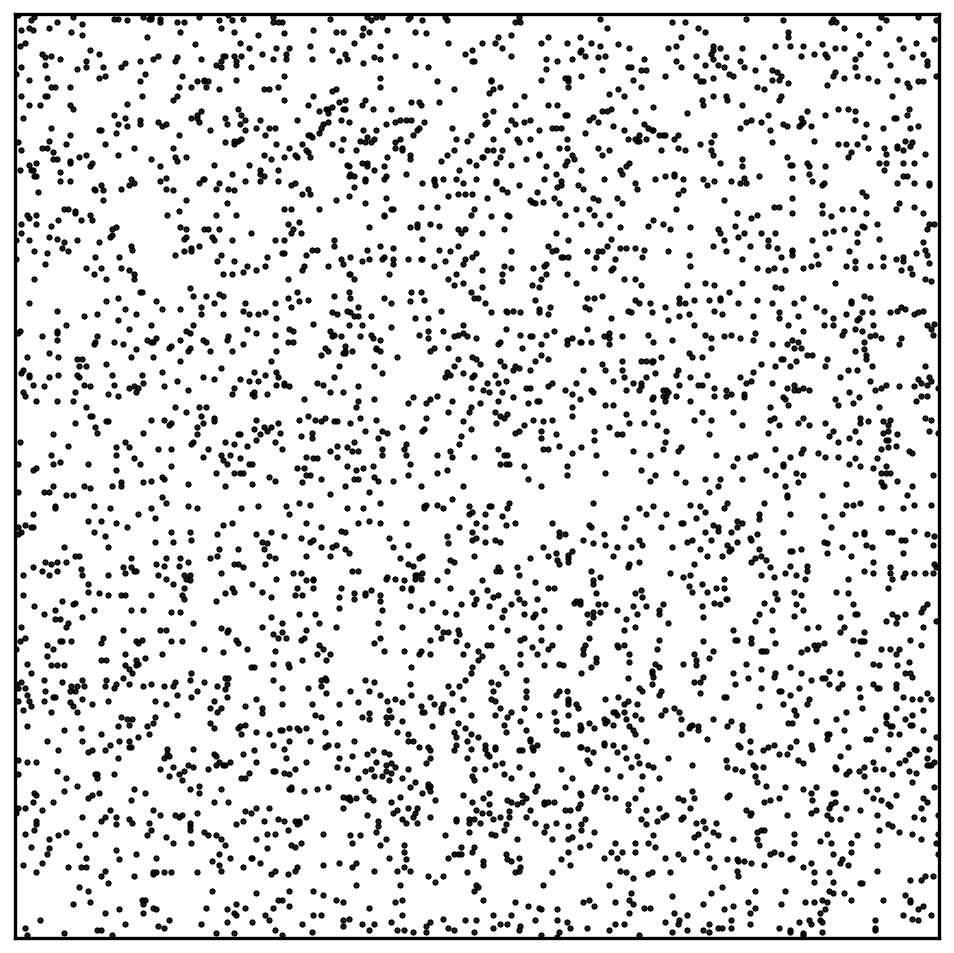}
        \caption{Uniform}
        \label{fig:mat_uniform}
    \end{subfigure}%
    \caption{Four of our generation methods: we can generate matrices following several distributions of nonzeros and that stress specific architectural features.}
    \vspace{-1em}
\end{figure}

Firstly, we test the impact of having an input with optimal spatial locality by adding a matrix comprised of \adria{a single dense row.} 
On certain workloads (i.e., \spmv, \spadd{}) this generates a regular access pattern that allows elements to be prefetched easily. Conversely, to test optimal temporal locality and simple branch prediction, we use a matrix comprised of \adria{a single dense column.} 
We then expand on this idea by including matrices that contain a cyclic pattern of nonzeros per row: this stresses the branch predictor in a controlled way, as the inner loops of \spmv{}, \spadd{}, and \spgemm{} operate on the elements of a single row at a time. 

To test the \gls{cpu}'s prefetchers, we include matrices that present elements in a strided pattern (\Cref{fig:mat_stride}) in which contiguous nonzeros of the matrix appear at $cache\_line\_size / \SI{4}{\byte}$ intervals.
To determine the impact of optimal spatial locality (\Cref{fig:mat_spatial}), we generate matrices with elements in clusters of $10$ elements, as it is an amount of nonzeros per row commonly found in literature  \cite{pooch1973survey,brayton1970some}. Conversely, to get optimal temporal locality (\Cref{fig:mat_temporal}), we generate matrices whose non-zeros always appear in the same columns.
Finally, we test the impact of several random distributions of nonzeros per row, typically found in real-world matrices such as scale-free graphs~\cite{bollobas2003directed, bollobas1998random,oskarsson2022scalable}. We achieve this by determining the number of nonzeros per row via uniform sampling of the inverse \gls{cdf} \cite{vogel2002computational} of the Gaussian, Exponential, and Uniform (\Cref{fig:mat_uniform}) distribution.
\Cref{tab:synth} summarizes the 9 categories of synthetic matrices we generate, along with the feature they outline.
The synthetic matrix generator is publicly available and can be found at \href{https://gitlab.bsc.es/fsgherzi/spchar-matrix-generator}{https://gitlab.bsc.es/fsgherzi/spchar-matrix-generator}. 
\begin{table}[!ht]
    \centering
    \caption{Synthetic matrices and their characteristics. The labels LOW, AVERAGE, HIGH refer to the a metric being below the first quartile (Q1), within the first and third quartile (Q1-Q3), and above the third quartile (Q3), respectively.}
    
  \resizebox{\textwidth}{!}{%
    \begin{tabular}{@{}lllll@{}}
    \toprule
        \textbf{Category} & \textbf{Temporal Locality} & \textbf{Spatial Locality} & \textbf{Row Imbalance} & \textbf{Branch Entropy}\\ \midrule
        Row & LOW & HIGH & HIGH & LOW \\ 
        Column & HIGH & HIGH & LOW & LOW \\ 
        Cyclic & LOW & LOW & LOW & AVERAGE \\ 
        Stride & LOW & HIGH & LOW & LOW \\ 
        Temporal & HIGH & LOW & LOW & LOW \\ 
        Spatial & LOW & HIGH & LOW & LOW \\ 
        Uniform & LOW & LOW & LOW & AVERAGE \\ 
        Exponential & AVERAGE & LOW & HIGH & LOW \\ 
        Normal & LOW & LOW & HIGH & AVERAGE \\ \bottomrule
    \end{tabular}}
    \label{tab:synth}
    \vspace{-1.5em}
\end{table} 


\subsection{Extracting static metrics from inputs}
We enrich the hardware counters' information with input-related metrics derived from analyzing the matrices. 
The fundamental operations of \spmv{}, \spadd{} and \spgemm{} concern operating on elements in a row-wise fashion (\Cref{alg:spmv,alg:spgemm,alg:spadd}), which inherently leads to high branch misprediction \cite{zhao2020exploring,kourtis2008optimizing} stemming from the fluctuation in the number of nonzero elements in the row. In this setting, \underline{Branch Entropy}  well encapsulates this feature and has been shown to correlate positively with branch miss rate \cite{yokota2008potentials}. 
\Cref{eq:branch_entropy} shows the formula used for computing branch entropy: 
\begin{align}\label{eq:branch_entropy}
    E &= - \sum_{i = 1}^N p(S_i)\log{p(S_i)} \\
    E_{max} &= -\sum_{i = 1}^N \frac{1}{N} \log{\frac{1}{N}} = -\log{\frac{1}{N}}
\end{align}
Where $S_i$ represents the length of a given branch (row size) and $p (S_i)$ the probability of encountering a loop of that size. Branch entropy is then normalized by $E_{max}$ to obtain a value between $0$ (no entropy, maximum prediction accuracy) and $1$ (maximum entropy, branch outcomes are not predictable).


\fra{Secondly, we categorize matrices with respect to the behavior they have on caches. For this, we determine their temporal and spatial locality.
A program or input has high temporal locality if, when a particular address is accessed, it is likely that it will be accessed again in the near future.
On the other hand, a program or input has high spatial locality if, when a particular address is accessed, addresses that are physically close in memory are likely to be accessed as well.}
Metrics that appropriately describe temporal and spatial locality are extremely complex to derive from the exploration of just inputs, as they heavily depend on the algorithm to which the inputs are applied. Upon examining the \spmv{} and \spgemm{} algorithms (\Cref{sec:motivation_and_background}) it can be noticed that, from a macroscopic perspective, they exhibit similar scan-and-lookup behavior, where elements on the left-hand side of the operation are streamed (scan) and used to index the right-hand side element (lookup). The left-hand side element does, by definition, exhibit optimal spatial locality and limited temporal locality, therefore, without loss of generality, we can focus on the locality pattern of just the right-hand side by extracting the list of indices being accessed. 
A well know instrument to determine the temporal locality of a set of addresses is the \underline{Reuse Distance} \cite{spirn1973program}, which has been shown to correlate positively to cache misses \cite{zhong2009program,keramidas2007cache}. 
Reuse distance is a metric that measures the number of unique memory addresses between two consecutive accesses to the same memory location. If an index is associated with a small reuse distance, the row accessed in the first memory access is likely to still be in the cache when the second memory access occurs. Conversely, the larger the reuse distance value for a given index, the higher the likelihood of the element being served from memory instead of caches, thus increasing the access time.
\fra{We determine spatial locality by employing the  \underline{Index Distance}, as defined by Fox et al. \cite{fox2008quantifying}. }
This method involves measuring the difference in index numbers between elements that are accessed subsequently by the algorithm. Lower values of average index distance imply that elements accessed subsequently are more likely to belong to the same cache line or to be prefetched together, whereas high values indicate a pattern that is more erratic and less cache/prefetcher friendly. 
Reuse and index distances are then transformed to $\log$-affinities (\Cref{eq:raba}) to clamp their values between $0$ (no affinity) and $1$ (maximum affinity) and to dampen the effect of extremal values \cite{fox2008quantifying}: accessing elements millions of indices apart has a similar cache behavior than accessing elements that are thousands of indices apart.

\begin{align}\label{eq:raba}
    \mbox{reuse\_affinity} &= \frac{1}{\log_{10}{(10 + \mbox{reuse\_distance})}} \\
    \mbox{index\_affinity} &= \frac{1}{\log_{10}{(10 + \mbox{index\_distance})}}
\end{align}

Multithreading performance is crucial in sparse algorithms. However, to best exploit the parallel capabilities of a \gls{cpu} the workload needs to be appropriately partitioned. The computation of \spmv, \spadd, and \spgemm{} can be trivially parallelized by partitioning the left-hand side matrix row-wise across multiple threads, as shown in \Cref{fig:matrix_partitioning}. The main drawback of this approach is that heavily imbalanced matrices (e.g. scale-free graphs with few big communities) hinder the multithread scalability of the row-wise partition scheme. To encapsulate this property, we present a metric called \underline{Thread Imbalance}\fra{, computed by determining the average deviation (as a ratio) between the number of nonzeros that each core receives using static partitioning and the optimal number of nonzeros per core (\Cref{eq:thread_imbalance})}.

\begin{figure}
    \centering
    \includegraphics[width=\linewidth]{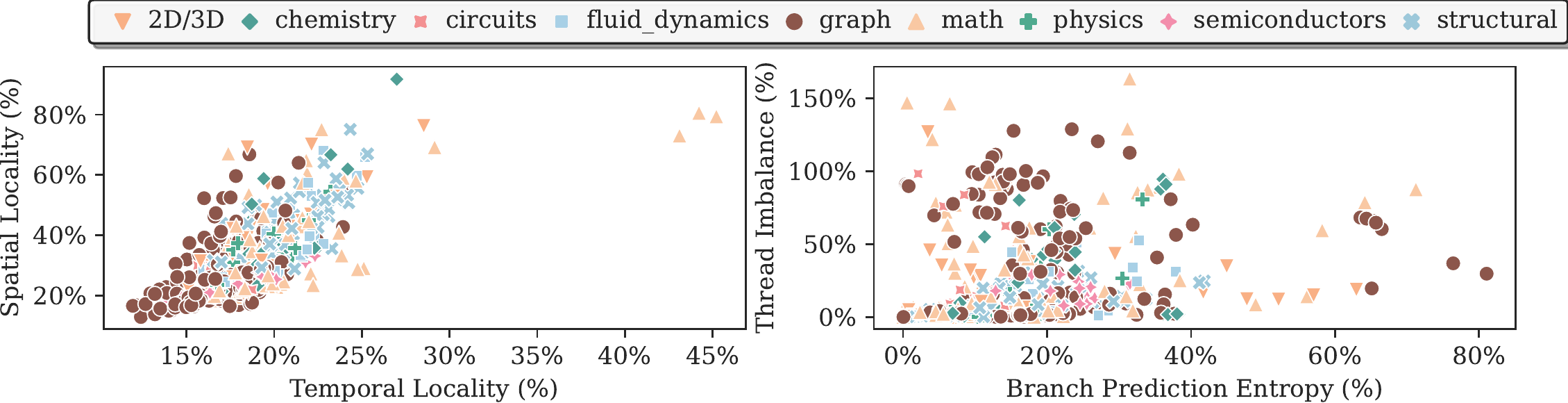}
    \caption{Temporal locality, Spatial locality, Branch entropy and Thread imbalance for our 9 matrix categories.} 
    \label{fig:matrix-categories}
\end{figure}

\begin{align}\label{eq:thread_imbalance}
    \mbox{thread\_imbalance} &= \frac{1}{T} \sum_{i = 1}^T \frac{\mid nnz_{assigned, i} - nnz_{ideal, i} \mid }{nnz_{ideal, i}} \\ 
    nnz_{ideal, i} &= \frac{nnz}{T} 
\end{align}

\begin{figure}
    \centering
    \begin{subfigure}{0.25\textwidth}
        \centering
        \includegraphics[width=0.8\textwidth]{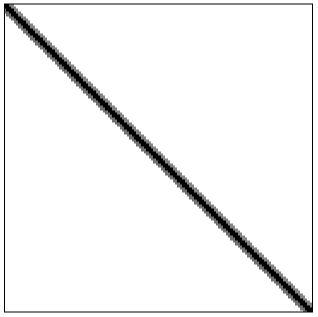}
        \caption{atmosmodd}
        \label{fig:mat_atmosmodd}
    \end{subfigure}%
    \begin{subfigure}{0.25\textwidth}
        \centering
        \includegraphics[width=0.8\textwidth]{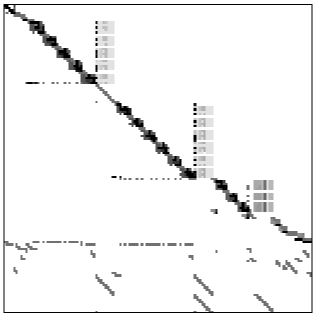}
        \caption{std1\_Jac2}
        \label{fig:mat_std}
    \end{subfigure}%
    \begin{subfigure}{0.5\textwidth}
        \centering
        \includegraphics[width=\textwidth]{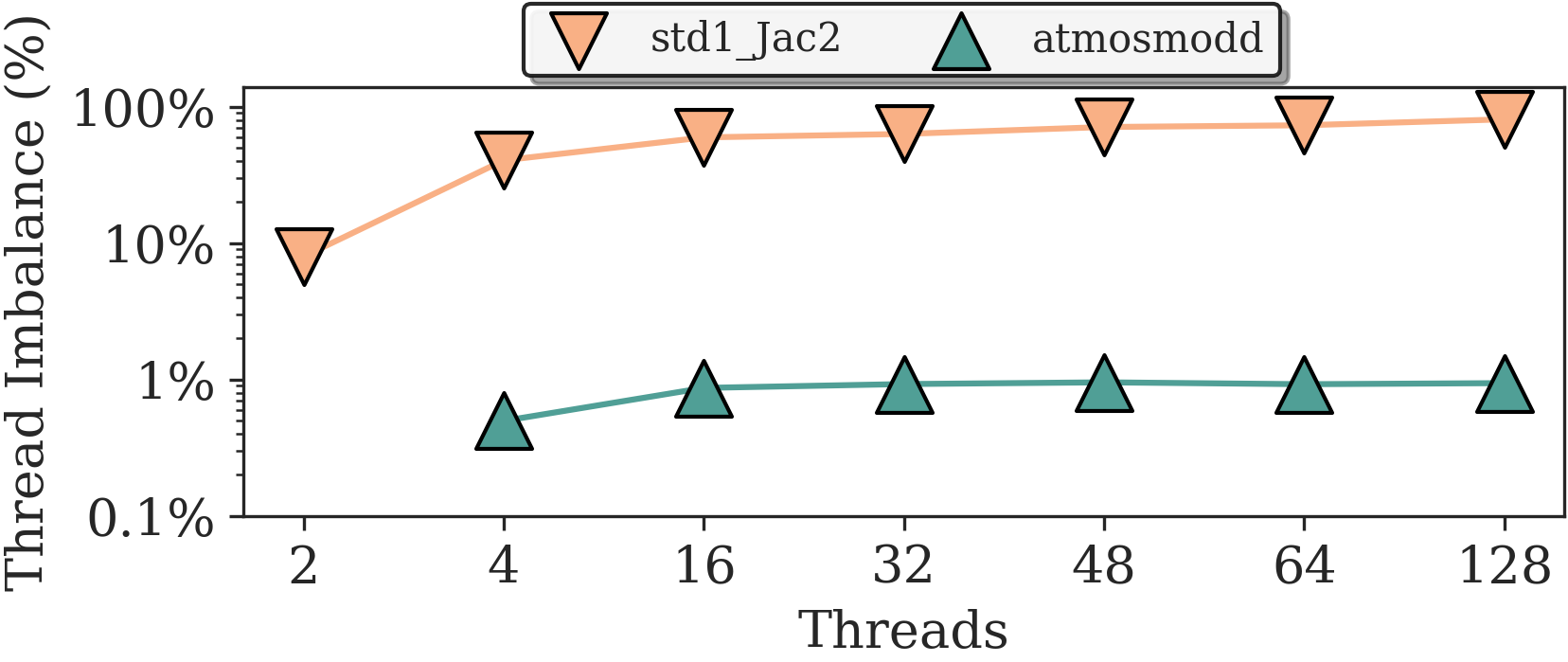}
        \caption{Thread Imbalance}
        \label{fig:thread_imbalance_vs}
    \end{subfigure}%
    \caption{Thread imbalance on two different matrices. (a) and (b) depict the sparse matrices as an adjacency matrix. \texttt{atmosmodd} \cite{suitesparse} exhibits a more consistent structure than \texttt{std1\_Jac2} \cite{suitesparse} which leads to it having orders of magnitude lower thread imbalance. We omit the value of thread imbalance on two threads for \texttt{atmosmodd} since it is $0$.}
    \label{fig:thread_imbalance}
\end{figure}

Where $T$ is the number of threads available in a system. For our analysis, we compute thread imbalance for $T \in [2, 4, 16, 32, 48, 64, 128]$. \Cref{fig:thread_imbalance} shows how \textit{thread imbalance} reacts to the increase of threads in a balanced (\Cref{fig:mat_atmosmodd}) and in an imbalanced (\Cref{fig:mat_std}) matrix.
\fra{Finally, in spite of several works \cite{6933066, xie2019ia, chen2020characterizing} taking into account the density of a matrix as a relevant metric for characterization, it rarely appeared in the list of most relevant features for a pair of \gls{cpu} and algorithm. We attribute this to the presence of the other metrics, which are much more fine-grained and are therefore able to be used more efficiently to encapsulate properties of the dataset.}
The code to compute the aforementioned metrics is publicly available and can be found at \href{https://gitlab.bsc.es/fsgherzi/spchar-matrix-analyzer}{https://gitlab.bsc.es/fsgherzi/spchar-matrix-analyzer}.

\Cref{fig:matrix-categories} displays the metrics for temporal locality, spatial locality, thread imbalance, and branch entropy grouped per matrix category. As expected, temporal and spatial locality are positively correlated ($\rho \approx 0.7$). \fra{Moreover, performing the ANOVA \cite{st1989analysis} statistical test between the matrix categories and the four aforementioned metrics yields p-values below $10^{-10}$. Hence we reject the null hypothesis and conclude that there is a statistically relevant correlation between categories and metrics.}

\subsection{Decision Trees for Extracting Relevant Architectural Insights} \label{sec:relevant_insights_extraction}
Determining what are the most relevant features of an architecture that is interacting with a problem requires knowledge of the problem, the inputs, and the architecture. However, it is not trivial to determine, for instance, which component of the \gls{cpu} interacts with which feature of the input or algorithm, as features like the presence of a certain branch predictor could affect, with similar impact, throughput, cycles spent stalling, vector unit utilization and many other factors. Conversely, those three axes cannot be analyzed by themselves in a vacuum: consider the case of a sufficiently small matrix that, in \spmv, has highly random accesses on the dense vector. Architectures that are able to fit the dense vector into cache entirely would yield much higher throughput than architectures that cannot. In \spadd, on the other hand, big or deeper caches are less impactful as the algorithm itself presents a much more regular access pattern to memory, compared to \spmv{} or \spgemm{}.

To extract the most relevant features of an architecture executing a certain algorithm on a given input we employ \fra{Tree-Based models} \cite{biau2016random,schapire2013explaining,hastie2009boosting,jia2013starchart}: Machine Learning algorithms that work by recursively partitioning the data set into smaller subsets based on the values of different features. In the context of regression (i.e., the objective is to infer a numerical variable), the algorithms select the feature that best separates the data into subsets with different characteristics, choosing the splitting attribute that minimizes the variance of the target variable. This process is repeated for each subset until the data can no longer be meaningfully partitioned. By the end of the process, the algorithm has built a tree-like structure with branches that correspond to different decisions based on the values of different features.

\fra{Within the family of tree-based models, there are various options available, with the typical trade-off being between interpretability and performance. In this setting, we employ decision trees, as interpretability is crucial in determining what are the most important features of an architecture. While more sophisticated tree-based models such as Random Forest \cite{biau2016random}, AdaBoost \cite{schapire2013explaining}, and Gradient Boosting Trees \cite{hastie2009boosting} might yield better performance, they are less interpretable by construction since they build collections of simpler trees and give them more importance the better they are able to predict a certain outcome. Consequently, the individual splitting attributes are not exposed, making it challenging to extract their most relevant features.} 

Within the scope of hardware architecture, decision trees can be used to analyze the performance of different architectures and identify the features that have the greatest impact on performance \cite{poe2008using}. This information can then be used to guide the design of new hardware architectures \cite{jia2013starchart} and optimize them for specific applications \cite{letras2021decision}.

We train our decision tree regressors over a slice of the whole dataset concerning a single algorithm (either \spmv{}, \spadd{} or \spgemm{}) and a single \gls{cpu} (either \postk{}, \kp{} or \graviton). \fra{ In this form, the dataset contains a row for each matrix and a column for each value of \gls{pmu} counters obtained by profiling the given algorithm, as well as the static metrics value associated with it. From this, we remove the \gls{pmu} counter values that have been used to compute target metrics, which, in the case of \gls{gflops}, are \texttt{ASE\_SPEC}, \texttt{VFP\_SPEC} and execution time.}
By setting \gls{gflops} as the target variable, we obtain the splitting attributes of the decision tree which, in turn, suggest a combination of architectural and most impactful input features. We then confirm the relevance of such attributes by comparing architectures: if an attribute is present among all of them it has a high likelihood of being a characteristic of the algorithm and conversely, if it is not, there exist architectural differences that justify a certain architecture not having a specific bottleneck. 

\section{Experimental evaluation}\label{sec:exp_res}
In this section, we present the evaluation of our methodology across the three \arm{} \glspl{cpu} and the matrix dataset. We first begin by assessing the capability of the decision tree models to encapsulate the properties of the dataset properly. We then move to an evaluation of the synthetic matrices to show how frontend and backend stalls relate to algorithms, inputs and ultimately architectural choices within the \glspl{cpu}. Finally, we present the analysis of the \spmv{}, \spadd{}, and \spgemm{} algorithms over the 600 real world matrices and propose key architectural and software improvements that can help overcome the bottlenecks characterizing these algorithms.

\subsection{Evaluating Characterization via Decision Trees}
\begin{figure}
    \centering
    \includegraphics[width=\linewidth]{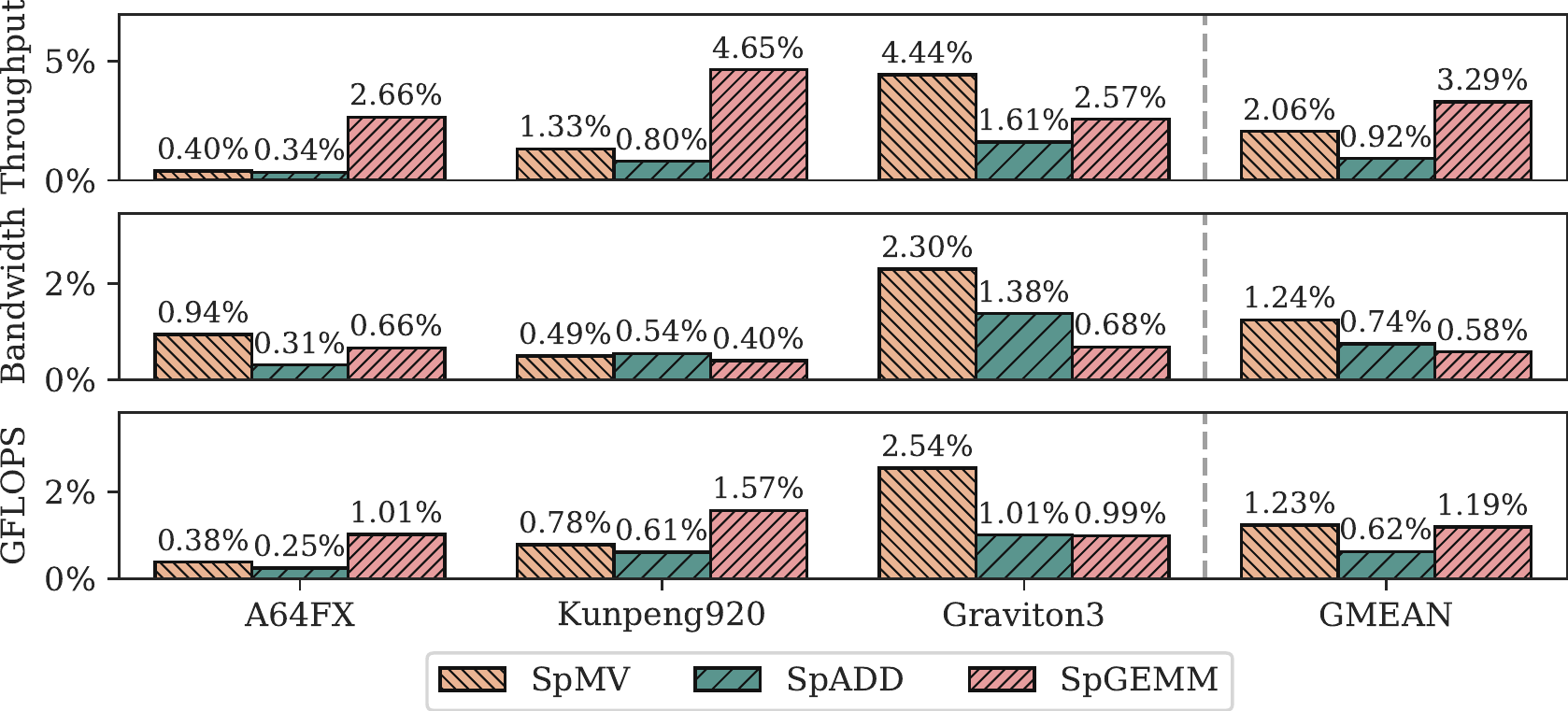}
    \caption{\gls{mape} of $10$-fold cross-validation applied to each \gls{cpu}.}
    \label{fig:mape_xfold}
\end{figure}
\glsreset{mape}
Before extracting relevant features from the decision trees, we need to confirm that our models are able to represent our dataset. To this end, we test the performance of our models through $K$-fold crossvalidation. $K$-fold crossvalidation works by partitioning the dataset into $K$ slices, training the model on $K - 1$ slices and testing on the leftover slice. This operation is then repeated $K$ times for all resulting permutations of training and testing splits. As target features for prediction, we choose GFLOPS, bandwidth, and throughput, expressed in terms of the number of iterations of the innermost loops for \spmv{}, \spadd{} and \spgemm{}. As outlined in \Cref{sec:relevant_insights_extraction}, we remove the \gls{pmu} counters that have been used to compute the target metrics from the dataset before training since this would make it trivial for the algorithm to predict the output perfectly. We use the \gls{mape} to estimate the goodness of our models, which presents a concise way to determine the deviations between predicted and actual values for regressions. In our testing, we opt for $K = 10$, i.e., performing the training and inference stages ten times over ten different partitions, as this value of $K$ is widely used in the literature~\cite{fushiki2011estimation,berrar2019cross,polat2007classification,banfield2006comparison}.

 \Cref{fig:mape_xfold} displays the average \gls{mape} of the $10$-fold crossvalidation per each target metric and each \gls{cpu}. Our models achieve low percentage error in predicting expected throughput, bandwidth, and GFLOPS for any combination of \gls{cpu} and algorithm: on average, the \gls{mape} is below $4\%$ indicating that the combination of our choice of performance counters and matrix metrics can accurately predict the target variables. \fra{Moreover, the median difference between the predicted and actual value is always below $0.001$ (i.e., $0.1\%)$ and coefficient of determination $R^2$ is always above $0.8$ indicating that our models are able to encapsulate the variance of the target features appropriately.}
 We observe that we are able to predict the performance of \postk{} and \kp{} better than \graviton{}, which is due to the fact that both Fujitsu and Huawei provide additional uncore counters in their optimization guide. At the time of writing this document, \gls{aws} does not provide documentation containing such \glspl{pmc}, therefore limiting our models to the counters provided by \arm.
  

\subsection{Evaluating Synthetic Matrices} \label{sec:eval_synth}
\begin{figure}
    \centering
    \includegraphics[width=\linewidth]{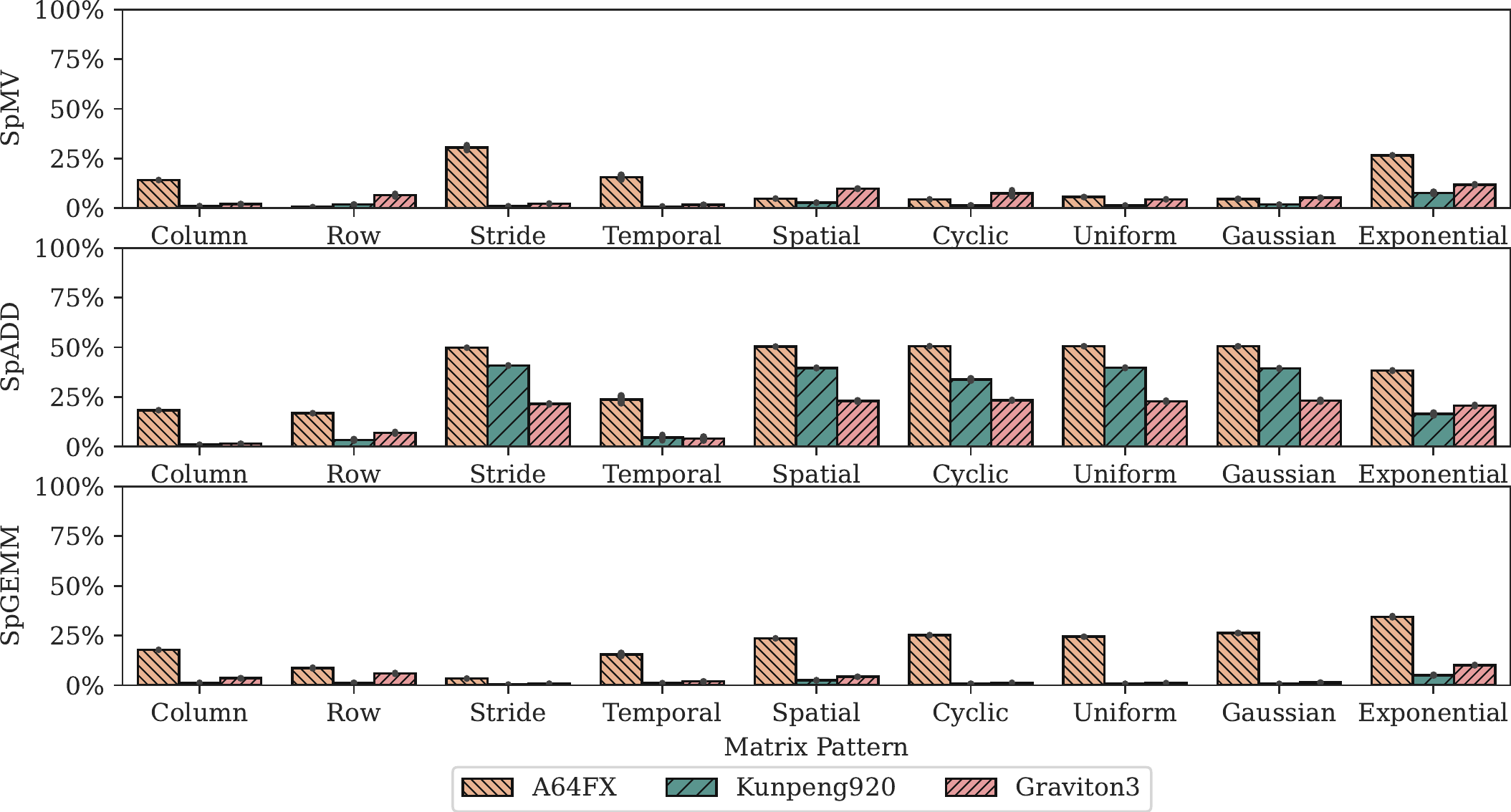}
    \caption{Percentage of Frontend Stalls over all cycles per algorithm and matrix pattern. Heavily branching codes like \spadd{} stall primarily in the frontend as a result of frequent bad speculation on \textit{data dependent} branches.}
    \label{fig:synth-fe-stall}
\end{figure}
We proceed with our analysis by showing how our synthetic matrices outline specific bottlenecks of \postk{}, \kp{}, and \graviton{}. \Cref{fig:synth-fe-stall} displays the percentage of cycles stalled in the frontend across our synthetic matrices. 
A \gls{cpu} experiences frontend stalls (i.e., instruction fetch and decode) when it cannot feed the execution units with new micro operations, which is commonly associated to instruction cache misses and bad speculation due to branch mispredictions. In our specific use case, kernels like \spmv{}, \spadd{}, and \spgemm{} are characterized by a limited amount of instructions to be executed at each loop iteration with no variations \textit{across loops}. On the other hand, as outlined in \Cref{sec:background}, the kernels in question have branches whose outcome depends on indirectly fetched values. We can therefore consider minimal the impact of instruction cache misses and attribute frontend stalls primarily to pipeline flushes resulting from bad speculation. 
On matrices that exhibit a more regular structure (Column, Row, Stride, Temporal), frontend stalls are low across the board, as a result of the more predictable row patterns. The Spatial and Cyclic matrices, albeit having a predictable structure in terms of memory accesses, do not have fixed row lengths and, as a result, exhibit similar patterns to the random matrices generated following a particular distribution. \spadd{} displays consistently high frontend stalls across all platforms and synthetic matrices, as expected from its highly branching code.
\spadd{} is comprised of two nested loops (\Cref{alg:spadd}), of which the inner depends on the length of the row of the two matrices. As a result, for each row, elements belonging to the rows of the two matrices are either summed (if their indices coincide) or copied into their appropriate slot (if the indices do not coincide). As this decision is data-dependent, it results in frequent bad speculations.
Regarding the difference in behavior between \glspl{cpu}, it can be noted that \postk{} has consistently higher frontend stalls, thus indicating that, in this platform, the cost of pipeline flushes resulting from bad speculation is much higher than in the competing platforms.

\begin{figure}
    \centering
    \includegraphics[width=\linewidth]{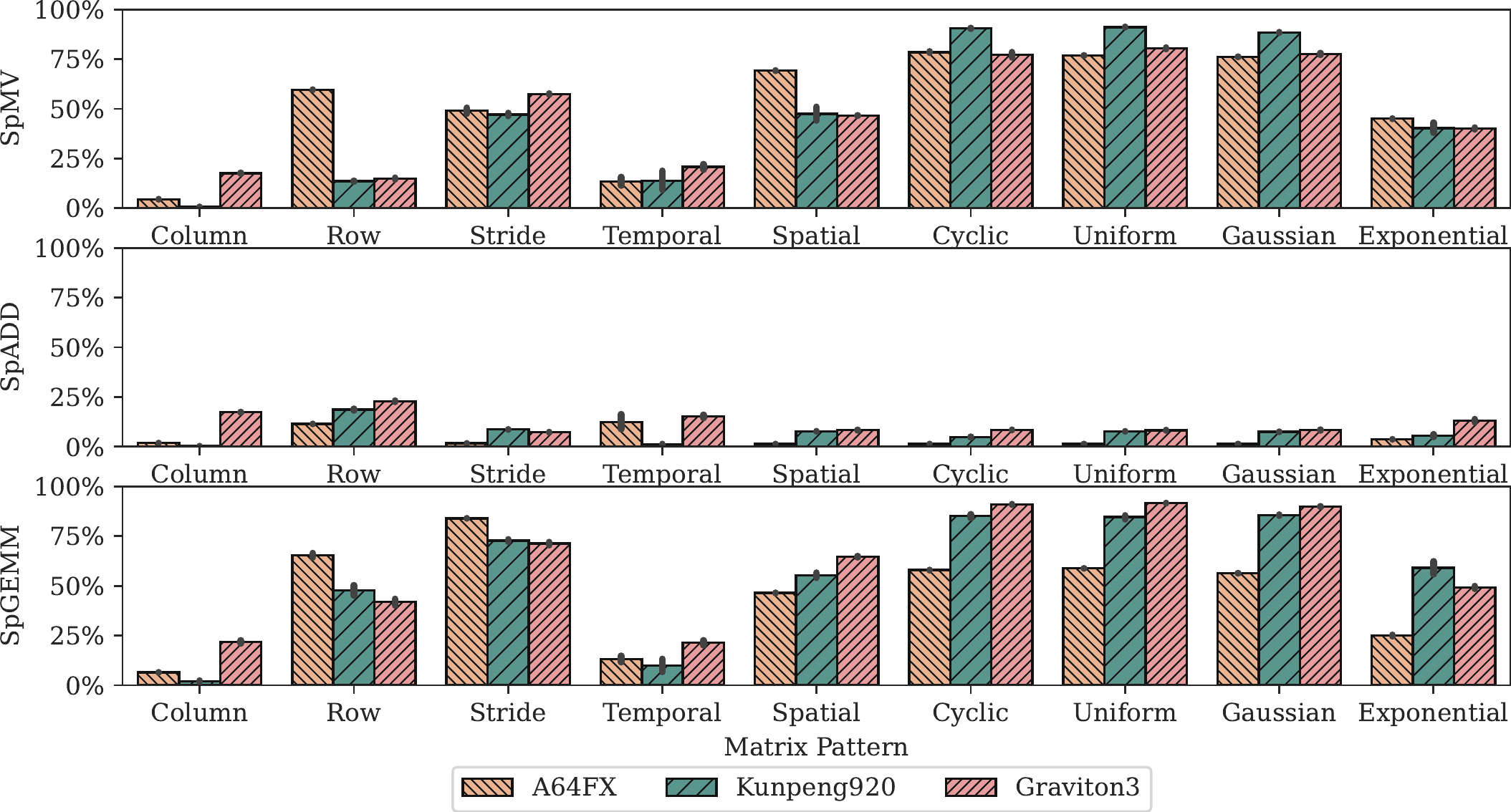}
    \caption{Percentage of Backend Stalls over all cycles per algorithm and matrix pattern. Memory heavy codes like \spmv{} and \spgemm{} incur in high backend stalls as a consequence of poor data locality and limited size of the \glspl{mshr}.}
    \label{fig:synth-be-stall}
\end{figure}
\Cref{fig:synth-be-stall} displays the percentage of cycles stalled in the backend for our synthetic matrices. The backend of the pipeline (i.e., instruction execution and memory operations) stalls when it needs to wait for the execution of arithmetic, vector, or memory instructions, which are particularly lengthy and therefore fill \glspl{mshr}, execution units and reorder buffer \adria{structures}.
Sparse codes have generally low operational intensity (i.e., their ratio of memory accesses per \adria{arithmetic} operation is high), therefore the majority of backend stalls can be attributed to long-latency memory operations, dictated by the latency of the memory subsystem and the number of \glspl{mshr}. This is further confirmed by the fact that codes that frequently gather data from indirect accesses (i.e., \spmv{} and \spgemm{}) stall heavily in the backend unless the pattern exhibits sufficient locality. As expected, \spmv{} is more sensitive to the matrix structure. Upon examining the code (\Cref{alg:spmv}), it is clear that for each output element, multiple indirect memory accesses have to be performed. This behavior quickly fills the \glspl{mshr}, unless the inputs expose high locality while referencing the dense vector. For instance, Column leads to repeated accesses to the same element of the dense vector, Row leads to a streaming access pattern on the dense vector, and Temporal also streams the dense vector but in bursts. 
\spgemm{} is similarly affected by this, as it shares the \textit{scan-and-lookup} behavior of \spmv{}. In contrast to \spmv{} however, the algorithm exhibits more \textit{intrinsic} locality as rows of both matrices need to be used to compute a single element. As a result, the magnitude of the amount of backend stalls is less dependent on the matrix structure and more dependent on the fact that \glspl{cpu} need to bring to caches entire rows/columns to compute single elements, which puts a lot of strain on the memory subsystem and trashes the caches. In contrast, \spadd{} does not exhibit a high amount of backend stalls, due to the fact that both the right-hand and the left-hand side matrices are streamed, resulting in a linear access pattern that exhibits optimal spatial locality.

\subsection{Extracting relevant features from decision trees}
As outlined in \Cref{sec:relevant_insights_extraction}, decision trees work by recursively partitioning the dataset based on the attribute value that minimizes the variance of the target feature. In this context, the amount by which the variance of the target is lowered is referred to as \textit{Gini Importance}. The byproduct of this is that attributes that display stronger relations with the target metric are chosen to be higher in the \textit{tree hierarchy}, whereas less related features appear lower and are given less importance. As a consequence, we can infer an ordering by retrieving the \textit{Gini Importance} of a given attribute, which, in turn, describes the relevance of a feature in predicting the target.
For visualization purposes, this relation is displayed in \Cref{fig:spmv-relevant,fig:spadd-relevant,fig:spgemm-relevant} by assigning to the rectangle containing hardware or input characteristics an area that is proportional to the importance of the feature in the induced decision tree. 
As we use our models for feature extraction and not for inference, we train our models on the entire dataset with the prediction target being GFLOPS.
\subsubsection{SpMV}
\begin{figure}
    \centering
    \includegraphics[width=\linewidth]{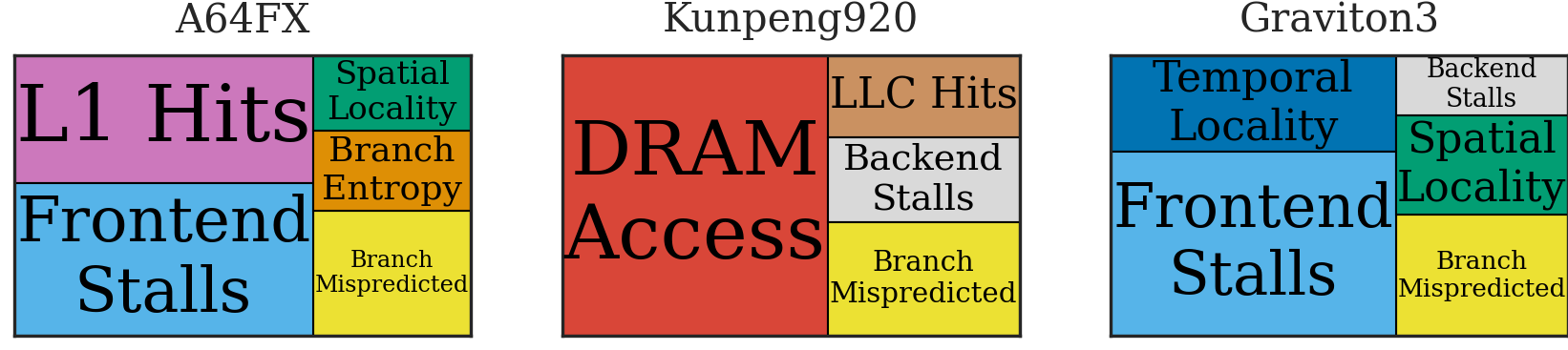}
    \caption{Most relevant input and hardware features for \spmv{}. \spmv{} is heavily dependent on the locality of the inputs and the amount of non-zeros per row, which determines the length of the inner loop.}
    \label{fig:spmv-relevant}
\end{figure}
\Cref{fig:spmv-relevant} displays the most relevant input and hardware features for each \gls{cpu}. From this, we draw three findings.

Firstly, in accordance with the state of the art in performance characterization for \spmv{} \cite{giannoula2022sparsep}, we determine that it is primarily bottlenecked by the latency of the memory system. As a result of the indirect access (\Cref{alg:spmv}), memory accesses are generally not prefetchable which makes the performance of \spmv{} to be highly reliant on the structure of the matrix (Spatial and Temporal locality attributes). 
\begin{figure}
    \centering
    \includegraphics[width=\linewidth]{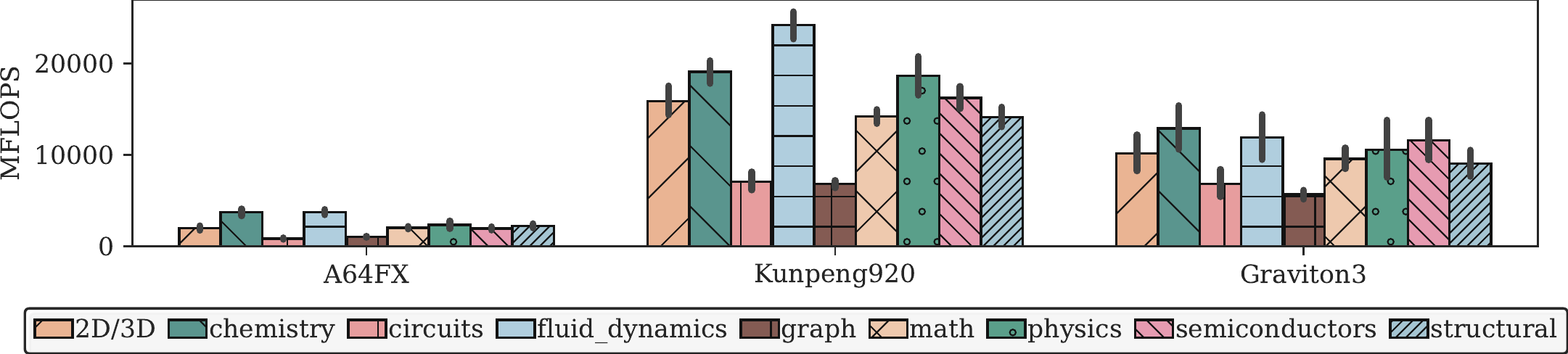}
    \caption{Performance of \spmv{} per \gls{cpu} and matrix category. The high number of memory channels of \kp{}, combined with the low latency of \gls{ddr}4 consistently contributes to higher performance.}
    \label{fig:performance_spmv}
\end{figure}
In accordance with this, the better the latency-under-load a system has, the better the performance.
To this end, \Cref{fig:performance_spmv} shows the performance in terms of MFLOPS for the three platforms, across matrix categories. \kp{} yields the best performance across the board, due to the lower memory latency of \gls{ddr}4 \cite{steiner2021exploration}, compared to \gls{hbm}2 in \postk{} and \gls{ddr}5 \graviton{} and the increased number of memory channels.  
\begin{figure}
    \centering
    \includegraphics[width=\linewidth]{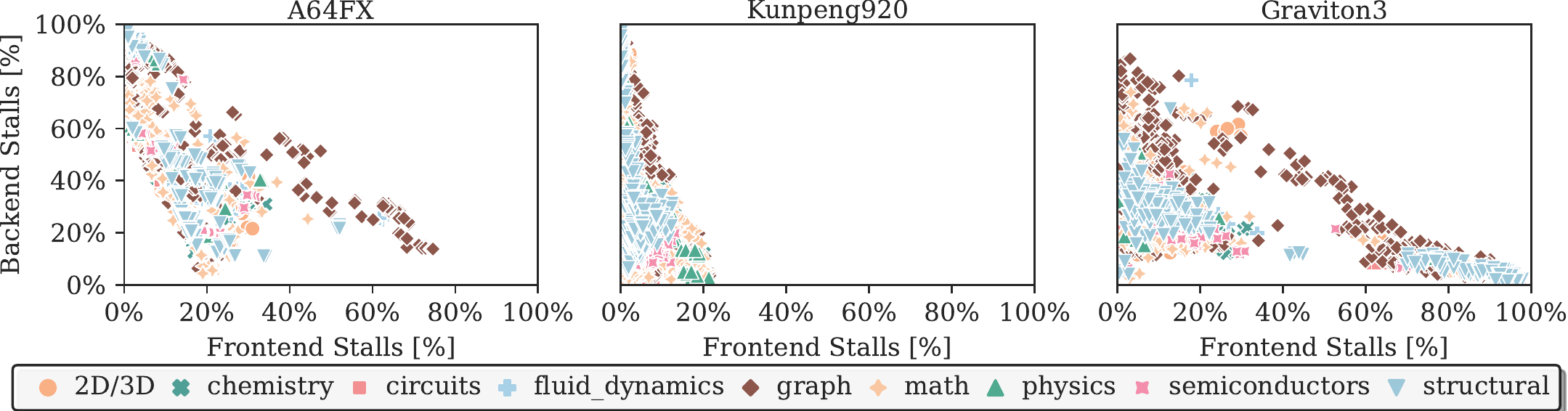}
    \caption{Percentage of cycles spent stalling in the frontend vs backend over all cycles for \spmv{}. Most of the stalls of \kp{} are skewed towards the backend, whereas for \postk{} and \graviton{} it leans more towards the frontend.}
    \label{fig:be-fe-stalls-spmv}
    \vspace{-1.2em}
\end{figure}

Secondly, we observe that frequent bad speculation has a high impact on performance and, on the three platforms we are examining, \postk{} and \graviton{} are the most affected. We investigate this further by examining the percentage of cycles stalled in the frontend and backend for each platform (\Cref{fig:be-fe-stalls-spmv}). For \graviton{} and \postk{}, more matrices incur in higher frontend stalls when computing \gls{spmv} compared to \kp{}. From this, we determine that, while there are architectural features that could help lower the time stalling in the backend (wider \gls{mshr}, lower memory latency), reducing frontend stalls from an architectural perspective is much trickier, as a more sophisticated branch predictor would still fail in predicting the outcome of branches when the decision is purely data driven. To overcome this issue, software designers can employ data structures that minimize the likelihood of branch mispredictions \cite{alappat2020performance} to occur and unrolling techniques \cite{gomez2021efficiently} to enable more efficient vectorization.

Thirdly, we determine that increasing the size of the private core caches and increasing the size of the \gls{mshr} can help overcome the dependency of performance from the inputs' locality pattern. Categories that generally exhibit good locality pattern (e.g., structural, semiconductors) see a decrease in the percentage of backend stalls on architectures with bigger caches (\graviton{}, \kp{}). In contrast, randomly structured matrices would benefit from the higher memory level parallelism enabled by wider \glspl{mshr}.

\subsubsection{SpGEMM}
\begin{figure}[t]
    \centering
    \includegraphics[width=\linewidth]{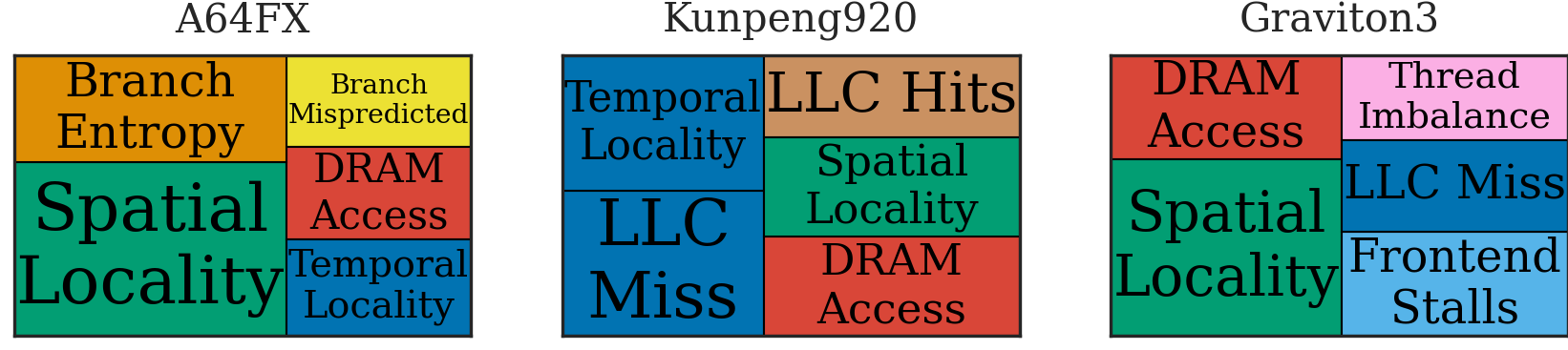}
    \caption{Most relevant model features for \spgemm{}: \spgemm{} greatly depends on the locality of the input, as a consequence of the frequent cache evictions that the algorithm inherently performs.}
    \label{fig:spgemm-relevant}
\end{figure}
\Cref{fig:spgemm-relevant} displays the most relevant model features for \spgemm{}. From this, we draw two findings.
Firstly, in spite of having a similar \textit{scan-and-lookup} memory access pattern as \spmv{}, the behavior from the cache perspective is much different, as \spgemm{} operates on entire rows and columns to produce an individual value of the output matrix.

\begin{figure}[t]
    \centering
    \includegraphics[width=\linewidth]{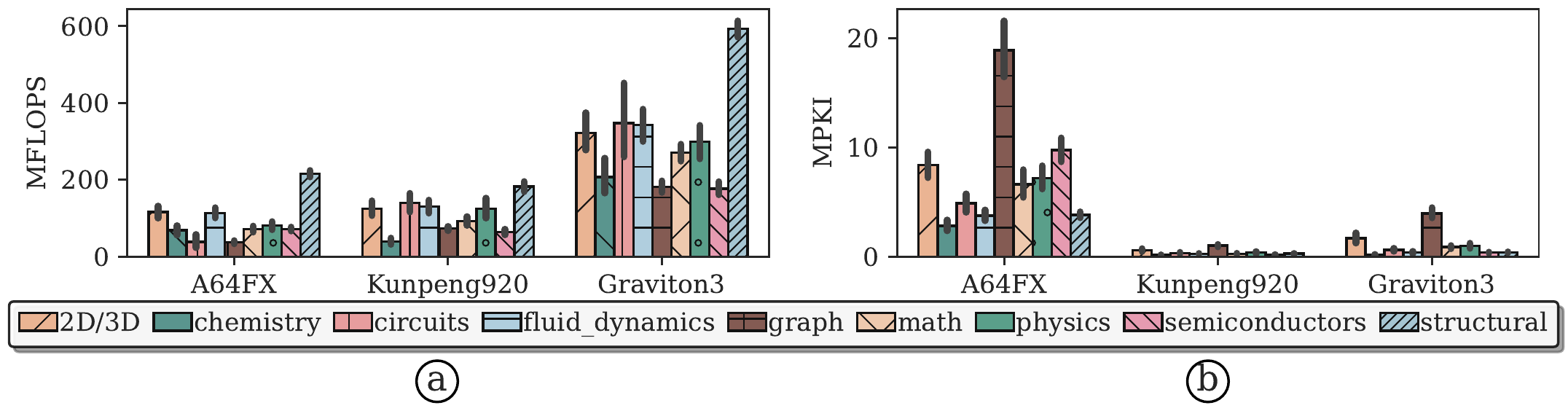}
    \caption{\textcircled{a} Performance of \spgemm{} per \gls{cpu} and matrix category. \graviton{} achieves better performances thanks to the combination of bigger caches and wider \glspl{mshr} to both exploit the locality of the matrices and having multiple memory requests in flight when needed. \textcircled{b} \gls{mpki} for \spgemm{}. Lower cache size and shallower cache hierarchy leads to \postk{} having an order of magnitude more \gls{mpki} compared to \graviton{} and \kp{}.}
    \label{fig:performance-mpki-spgemm}
\end{figure}
\begin{figure}
    \centering
    \includegraphics[width=\linewidth]{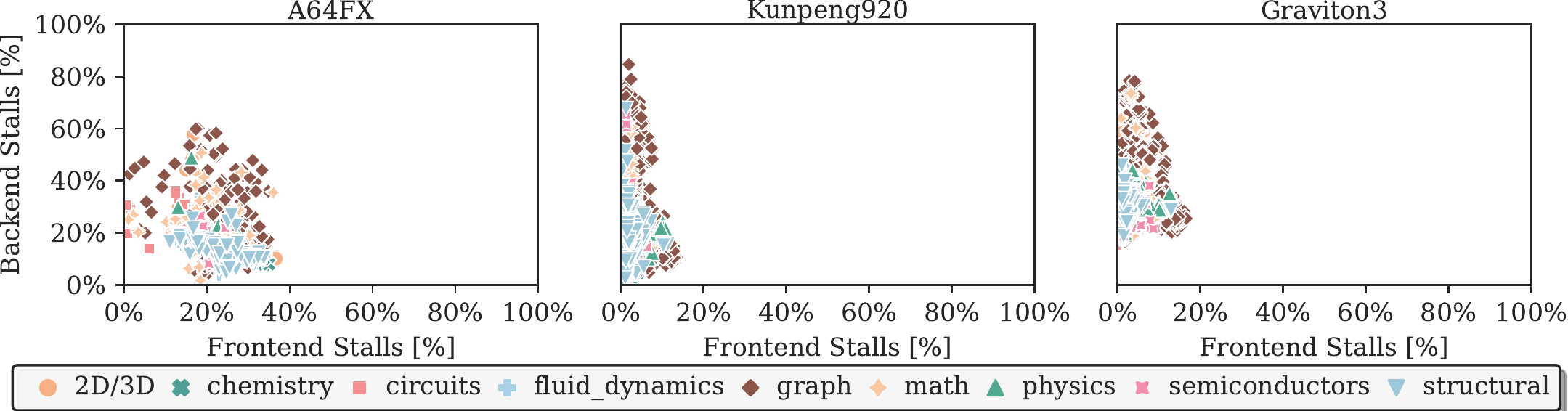}
    \caption{Percentage of cycles spent stalling in the frontend vs backend over all cycles for \spgemm{}. Differently from \spmv{}, \glspl{cpu} executing \spgemm{} stall more frequently waiting for the memory system as a result of inherently poor reuse patterns of the algorithm.}
    \label{fig:be-fe-stalls-spgemm}
\end{figure}

As a result, caches have a higher likelihood of being polluted by values that are rarely reused after first touch, thus making evictions more frequent. Systems with shallower cache hierarchies such as \postk{} are more affected by this behavior, as shown in \Cref{fig:performance-mpki-spgemm} \textcircled{b}, having an order of magnitude more \gls{mpki}.
Moreover, this puts a heavier strain on the memory system (\Cref{fig:be-fe-stalls-spgemm}) as new values need to be fed continuously to keep the \glspl{alu} active, which results in systems with either lower memory access latency (\kp{}) or bigger \glspl{mshr} (\graviton{}) to achieve better performance, as shown in \Cref{fig:performance-mpki-spgemm} \textcircled{a}.

Secondly, matrix domain has an impact on the performance of \spgemm{}. This is not only a byproduct of being sensitive to the locality of the inputs, but also stems from the fact that different matrix categories have different distributions of number of nonzeros across rows. Albeit not as branch-heavy as \spmv{}, a branch misprediction in \spgemm{} still incurs in a high penalty as a result of having to discard the result of plenty of memory instructions when a misprediction occurs.

\subsubsection{SpADD}
\begin{figure}
    \centering
    \includegraphics[width=\linewidth]{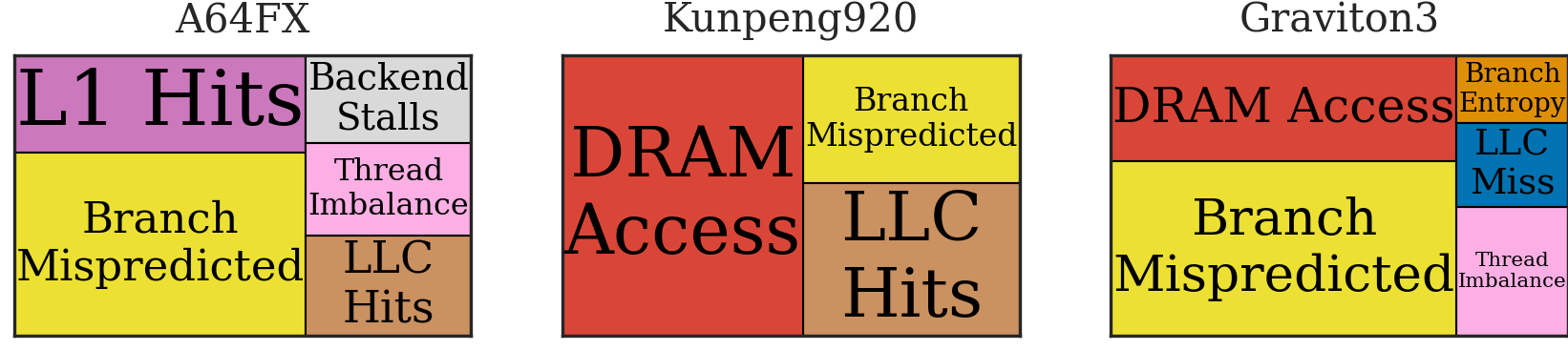}
    \vspace{-15pt}
    \caption{Most relevant model features for \spadd{}: Non predictable branches are frequent in \spadd{}, which translates to it being a relevant feature for our models.}
    \label{fig:spadd-relevant}
\end{figure}
\Cref{fig:spadd-relevant} shows the most relevant hardware and input features for each \gls{cpu}. From this, we draw three findings. 
Firstly, performance in \spadd{} is heavily dependent on the pipeline flush overhead that results from a branch misprediction, which is in line with our findings shown in \Cref{sec:eval_synth} for the synthetic matrices. As detailed in \Cref{sec:background_spadd}, this kernel stresses the branch predictors heavily due to the continuous \textit{sum-or-merge} operation that occurs for each row. 
\begin{figure}
    \centering
    \includegraphics[width=\linewidth]{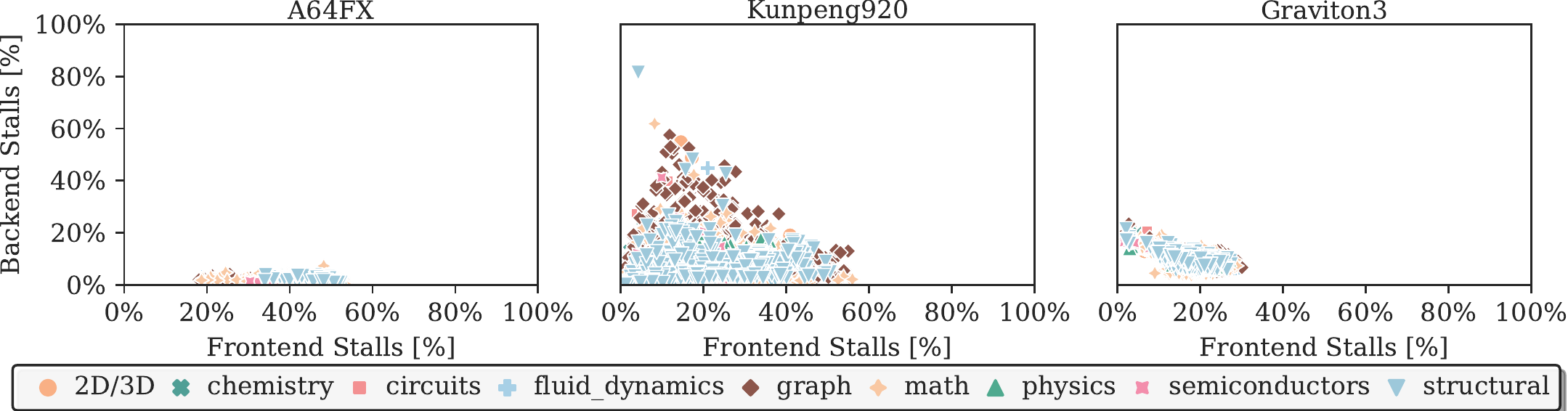}
    \caption{Percentage of cycles spent stalling in the frontend vs backend over all cycles for \spadd{}. High memory bandwidth and advanced prefetchers limit the time spent stalling in the backend, therefore leaning more towards stalling for branch mispredictions.}
    \label{fig:be-fe-stalls-spadd}
\end{figure}
This is only further stressed by the fact that even matrices that have similar structures will incur in this, if the nonzeros in each row are not found in similar positions.

\begin{figure}
    \centering
    \includegraphics[width=\linewidth]{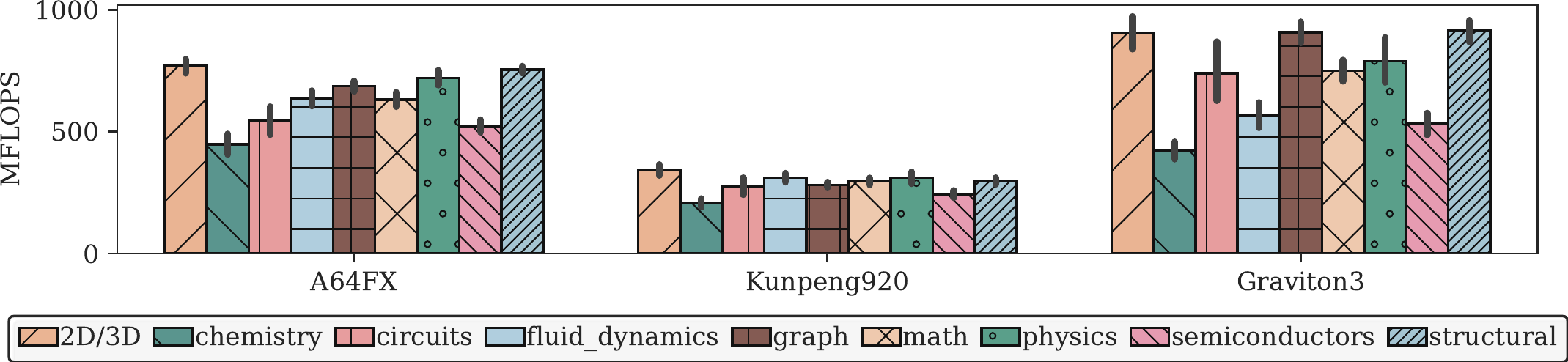}
    \caption{Performance of \spadd{} per \gls{cpu} and matrix category. Higher memory bandwidth (\postk) or better prefetchers (\graviton) contributes to higher performance.}
    \label{fig:performance-spadd}
\end{figure}

Secondly, stemming from the fact that memory accesses of \spadd{} are easily prefetchable, platforms with more aggressive prefetchers and higher memory bandwidth achieve better performance. This is confirmed by \Cref{fig:be-fe-stalls-spadd} and \Cref{fig:performance-spadd}, where \postk{} and \graviton{} are able to pull up better overall performance compared to \kp{}, as a result of better prefetcher behaviors and higher memory bandwidth. 

Thirdly, \spadd{} is less dependent on the matrix category than \spmv{} and \spgemm{}. The memory accesses performed by this kernel are contiguous, as the matrices are streamed linearly from main memory. As a result, the in-row locality, outlined by the attributes LLC Hits/Miss, is more relevant than the overall locality that would stem from a specific matrix category.

\subsection{Optimizing software and hardware architectures for sparse computation}\label{sec:optimizing}
Having discussed what are the most relevant architectures and input features for \glspl{cpu} performing sparse computation, we now propose suggestions to hardware and software architects to implement efficient architectures and algorithms for sparse computation.

\noindent \textbf{\spmv{}.} If the inputs exhibit low locality, the main performance benefits can be obtained by increasing the capability of the architecture to have multiple outstanding memory operations. This can be performed in two orthogonal ways, by increasing the size of the \glspl{mshr} per core and increasing the size of the \gls{rob} or by reducing the \textit{load-to-use} latency. As caches have a low impact in this scenario, using memory technologies and architectures with low latency will prove to be beneficial. 
Moreover, as the input matrix is streamed and accesses have optimal temporal locality, performance benefits in traditional architectures could be obtained by pairing a high bandwidth memory for streaming and a low latency memory for random accesses in a heterogeneous memory configuration. In turn, this will additionally help reduce the pressure on the memory controllers, as streaming accesses will be relegated to an interface that has, by design, more memory channels.
Based on this, \gls{pim}, \glspl{fpga}, and decoupled access-execute architectures \cite{siracusa2023tmu} could be efficiently employed for this kernel. 
The higher the locality of the input, the lower the impact of the memory subsystem. As a result, the overhead of branch misprediction becomes more relevant. 
Branch mispredictions stem from the fact that the number of nonzeros in a row of a sparse matrix have high variance. Hence, to overcome this limitation, software architects can employ data representation formats that uniform matrices' rows in fixed size chunks, therefore making branches have a fixed length and rendering them more predictable (\gls{ell} and variants \cite{zheng2014biell,wang2020pellr,giannoula2022sparsep}).

\noindent \textbf{\spgemm{}.} Similarly to \spmv{},  \spgemm{} exhibits a \textit{scan-and-lookup} behaviour. As a result, the recommendations that apply to \spmv{} also apply to \spgemm{}.
Additionally, the performance of \spgemm{} is greatly dependent on the locality of the inputs. Low locality implies continuous cache evictions resulting from fetching entire rows/columns of the matrices to perform the computation of single output values. Therefore, the major performance improvements would be obtained by using large caches or forcing more input locality via matrix reordering \cite{liiv2010seriation}, and block-based partitioning mechanisms.

\noindent \textbf{\spadd{}.} \spadd{} is much less dependent on the matrix structure, having both input matrices being streamed to perform the computation. As a result, it benefits greatly from prefetching and having a high bandwidth memory subsystem. 
The heavy branching behavior can be overcome by using data structures that regularize the branching pattern. In this regard, formats that would benefit \spadd{} the most are 2D block-based \cite{giannoula2022sparsep,ccatalyurek2010two,pelt2014medium}, combining coalesced memory accesses at the row-level with predictable branching behavior.

\subsection{Summary and applicability of the SpChar methodology}
In this section, we have introduced a new methodology apt to determine the features that are most impactful for sparse computation. This is achieved by extracting insights from static input metrics and \gls{pmu} counters. We have then analyzed these insights to provide suggestions to software and hardware architects seeking to optimize sparse computation. 

The next step would be to prove that the optimizations we propose are effective and to provide quantitative data on the potential improvements. To this end, \Cref{sec:ch_loop} uses the \spmv{} kernel as a use case to explore the optimization proposed, and relates them to matrices that display the specific structures we have outlined to be problematic for sparse computation.

As stated in \Cref{sec:optimizing}, key factors influencing \spmv{} performance are the memory subsystem's latency for indirect access, its bandwidth for streaming accesses, and the efficiency of the frontend in managing the cost of incorrect speculation. Therefore, architectural enhancements for \spmv{} should focus on enabling the architecture to feed data to the core with reduced latency and minimizing the impact of frontend stalls. These optimizations can be translated to increasing the size of \glspl{mshr} and \gls{rob}, as well as employing prefetchers. Additionally, memory interfaces should be selected based on the access pattern they primarily service. We explore these parameters in \Cref{sec:ch_loop}.

\section{Optimizing SpMV through SpChar: the Characterization Loop}\label{sec:ch_loop}
\glsunset{dram}
To complete the characterization loop that SpChar provides, we now implement the suggestions we have made in \Cref{sec:optimizing}. To this end, we analyze a common use case of optimizing an architecture for the execution of the \spmv{} kernel.

\begin{table}[!ht] 
    \centering
    \caption{Summary of the features of the simulated \gls{cpu}}
    \begin{tabular}{@{}ll@{}}
    \toprule
        \textbf{Architecture} & ARM v8.2-a \\ 
        \textbf{Cores} & 16 Out Of Order Cores at 2.4GHz \\
        \textbf{Vector units} & 2 $\times{}$ 256-bit SVE \\
        \textbf{L1D, L1I per core} & \SI{64}{\kilo\byte}, 4-way, 2 cycle data access \\ 
        \textbf{Private L2 per core} & \SI{1}{\mega\byte}, 8-way, 4 cycle data access \\ 
        \textbf{Shared LLC} & 16 $\times{}$ \SI{1}{\mega\byte}, 16-way, 10 cycle data access \\ 
        \midrule
        \textbf{Network-on-chip} & $8\times8$ 2D mesh with AMBA 5 CHI, \\ &  1-cycle route and link latency\\
        \midrule
        \textbf{Memory technology} & \gls{ddr}5, \gls{hbm}2e \\ 
        \textbf{Memory channels} & 4, 8 \\ 
        \textbf{\gls{numa} Nodes} & 2 \\
        \textbf{Peak bandwidth} & \SI{452.8}{\giga\byte / \second} \\ \midrule
        \textbf{Operating System} & Ubuntu 22.04 \\
        \textbf{Compiler} & \texttt{g++ 11.3.1}  \\ 
        \bottomrule
    \end{tabular}
    \label{tab:simcpu}
\end{table}

\begin{table}[!ht] 
    \centering
    \caption{Summary of the explored parameters. Values in bold indicate the default.}
    \begin{tabular}{@{}ll@{}}
    \toprule
        \textbf{L1D \glspl{mshr}} & \textbf{8}, 16, 32 \\
        \textbf{L1D Prefetcher} & \textbf{None}, Stride-degree 2 \\  
        \textbf{L2 \glspl{mshr}} & \textbf{16}, 32, 64 \\
        \textbf{L2 Prefetcher} & \textbf{None}, Best Offset \cite{michaud2016best} \\  
        \textbf{LLC \glspl{mshr}} & \textbf{32}, 64, 128 \\
        \textbf{ROB size} & \textbf{224}, 554 \\
        \midrule
        \textbf{Addr. mapping for Memory Device} & \textbf{RoCoRaBaCh}, RoRaBaChCo\\
        \textbf{Memory Row Policy} & \textbf{open}, close \\
        \midrule
        \textbf{Vectorization} & \textbf{Compiler},  \gls{sve} Intrinsics \\
        \textbf{Sparse Matrix Location} & \textbf{\gls{ddr}5},  \gls{hbm}2e \\
        \textbf{Dense Vector Location} & \textbf{\gls{ddr}5}, \gls{hbm}2e \\
        \bottomrule
    \end{tabular}
  
    \label{tab:simparams}
\end{table}
\begin{table}[!ht] 
    \centering
    \caption{Summary of the presented configurations.}
    \begin{tabular}{@{}ll@{}}
    \toprule
    \textbf{ID} & \textbf{Explored Parameters}                                  \\ 
    \midrule
    \textbf{C1} & Defaults (bold parameters in \Cref{tab:simparams})            \\
    \textbf{C2} & \textbf{C1} + L1D, L2 prefetchers                             \\  
    \textbf{C3} & \textbf{C2} + 32, 64, 128 \glspl{mshr} for L1D, L2, LLC respectively       \\
    \textbf{C4} & \textbf{C3} + 554 \gls{rob} entries                                      \\ 
    \textbf{C5} & \textbf{C4} + Dense Vector in \gls{ddr}5 \& Sparse Matrix in \gls{hbm}2e \\ 
                & \gls{ddr}5 uses closed row policy and RoCoRaBaCh Addr. mapping.          \\ 
                & \gls{hbm}2e uses open row policy and RoRaBaChCo Addr. mapping.           \\
    \bottomrule
    \end{tabular}
    \label{tab:simconfigs}
\end{table}

We use the gem5 simulator \cite{lowe2020gem5} to simulate a 16-core \arm{} SoC, based on the Neoverse N1 architecture. The \gls{noc} is designed to support two \gls{numa} nodes. Each \gls{numa} node is associated with a specific type of memory technology – one with four channels of \gls{ddr}5 memory and the other with 8 channels of \gls{hbm}2e.
\gls{cpu} features are displayed in \Cref{tab:simcpu}. 
Starting from this configuration, we consider the performance bottlenecks that affect it and incrementally alleviate them by changing a single architectural feature (\Cref{tab:simparams}) at a time, thus allowing us to determine the individual contribution of each feature to the performance increase. We repeat this loop of choosing an architectural feature, applying it, benchmarking, and analyzing the impact on performance until we exhaust the exploration space. Due to space constraints, we present five notable configurations, outlined in \Cref{tab:simconfigs}.

The first architectural optimization we test is the inclusion of private cache prefetchers (\textbf{C2}), which aid in streaming the sparse matrix. Configuration \textbf{C3} tackles the issue of tolerating cache misses by increasing the \glspl{mshr}' size, thus easing the burden of random accesses to the dense vector. Configuration \textbf{C4} builds upon \textbf{C3} by increasing the size of the reorder buffer, hence increasing the number of in-flight memory requests that the \glspl{cpu} can have. Finally, \textbf{C5} tackles the issue of having streaming-like accesses on the sparse matrix and random accesses on the dense vector. To this end, we place the data structures that are accessed in a streaming fashion in \gls{hbm}2e on \gls{numa} Node 1 and the ones that are accessed randomly on \gls{ddr}5 on \gls{numa} Node 0 using the \texttt{numa\_alloc\_onnode} function from \texttt{libnuma}. Additionally, we tweak the address mapping and row policy of such interfaces to optimize the accesses they ought to serve. In this setting, \gls{hbm}2e operates under the Open Row policy (i.e., keep the row buffer open after access) with the RoRaBaChCo (row, rank, bank, channel, column) address mapping. This configuration capitalizes on linear accesses as contiguous addresses are more likely to hit a bank that is already precharged.
Conversly, \gls{ddr}5 operates under Closed Row policy (i.e, banks are precharged after every access) and the RoCoRaBaCh (row, column, rank, bank, channel) address mapping. Random accesses are not expected to hit a row buffer that is already open, hence, by closing it after access, bank precharge times are not on the critical path.

\begin{table}[!ht]
    \centering
    \caption{Matrices used in the evaluation and their characteristics, sorted by number of non-zero entries.  The labels LOW, AVERAGE, HIGH refer to the a metric being below the first quartile (Q1), within the first and third quartile (Q1-Q3), and above the third quartile (Q3), respectively.}
  \resizebox{\textwidth}{!}{%
    \begin{tabular}{@{}llllllll@{}}
    \toprule
        \textbf{Name} & \textbf{ID} & \textbf{Rows} & \textbf{NNZ} & \textbf{Temporal Loc.} & \textbf{Spatial Loc.} & \textbf{Row Imbalance} & \textbf{Branch Entropy}\\ \midrule
        std1\_Jac2 & M1 & 22K & 1.2M & HIGH & HIGH & AVERAGE & AVERAGE \\
        test1 & M2 & 392.9K & 13.0M & HIGH & LOW & AVERAGE & AVERAGE\\
        nd12k & M3 & 36.0K & 14.2M & HIGH & HIGH & HIGH & AVERAGE\\
        TSOPF\_RS\_b2383 & M4 & 38.1K & 16.2M  & LOW & LOW & HIGH & HIGH \\
        human\_gene2 & M5 & 14.3K & 18.1M  & HIGH & HIGH & HIGH & AVERAGE\\
        Transport & M6 & 1602.1K & 23.5M  & LOW & AVERAGE & LOW  & LOW \\
        nd24k & M7 & 72.0K & 28.7M & HIGH & HIGH & HIGH & AVERAGE \\
        mawi\_201512012345 & M8 & 18571.2K & 38.0M  & AVERAGE &  HIGH & LOW & HIGH\\
        spal\_004 & M9 & 10.2K & 46.2M & HIGH & AVERAGE & HIGH & LOW \\
        \bottomrule
    \end{tabular}}
    \label{tab:simmatrices}
\end{table}

To limit exploration and simulation times, we test our configurations on a subset of 9 matrices from our main real-world dataset. To this end, we choose the top 2 matrices that had the highest (for branch entropy and thread imbalance) or lowest (index and reuse affinity) values of the four static metrics that we have presented in \Cref{sec:methodology}, plus a small matrix (\textbf{M1}) to outline the effect of being able to cache the data structures involved in the computation entirely. The selected matrices are detailed in \Cref{tab:simmatrices}.

\begin{figure}[t]
    \centering
    \includegraphics[width=\textwidth]{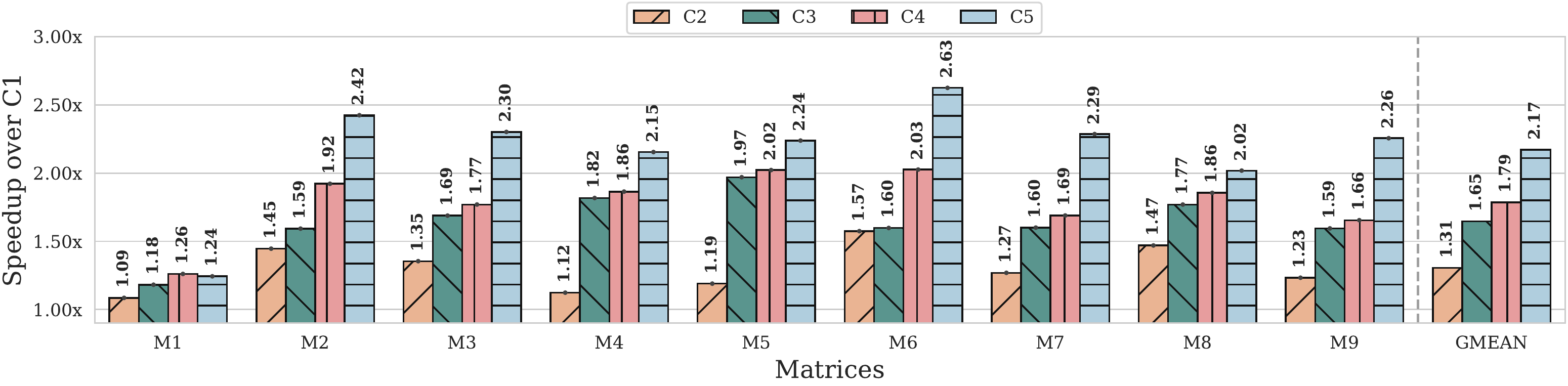}
    \caption{Speedup with respect to default configuration (\textbf{C1}). Having dedicated memory interfaces for streaming and random accesses greatly improves execution times (\textbf{C3}).}
    \label{fig:sim_speedup}
\end{figure}

\Cref{fig:sim_speedup} shows the speedup of each configuration normalized to \textbf{C1}.
From this, we draw three conclusions. 

Firstly, adding private cache prefetchers (\textbf{C2}) reduces the execution time by an average 31\%. While the accesses on the dense vector are generally not predictable, the 80\% of the data involved in the computation (i.e., output, nonzero, row pointer, column index vectors) is accessed linearly, resulting in prefetchers being able to gather data from memory beforehand effectively. On the other hand, this category of prefetchers fail in proactively retrieving the next datum from the dense vector, as they are unable to infer access patterns in indirect, random accesses. While prefetchers like \gls{imp} \cite{yu2015imp} might yield better performance, they would only directly benefit kernels with this kind of access pattern and not improve applications that use this kernel along with others \cite{parravicini2021reduced, parravicini2021scaling,sgherzi2022eigen}.

\begin{figure}[t]
    \centering
    \includegraphics[width=\textwidth]{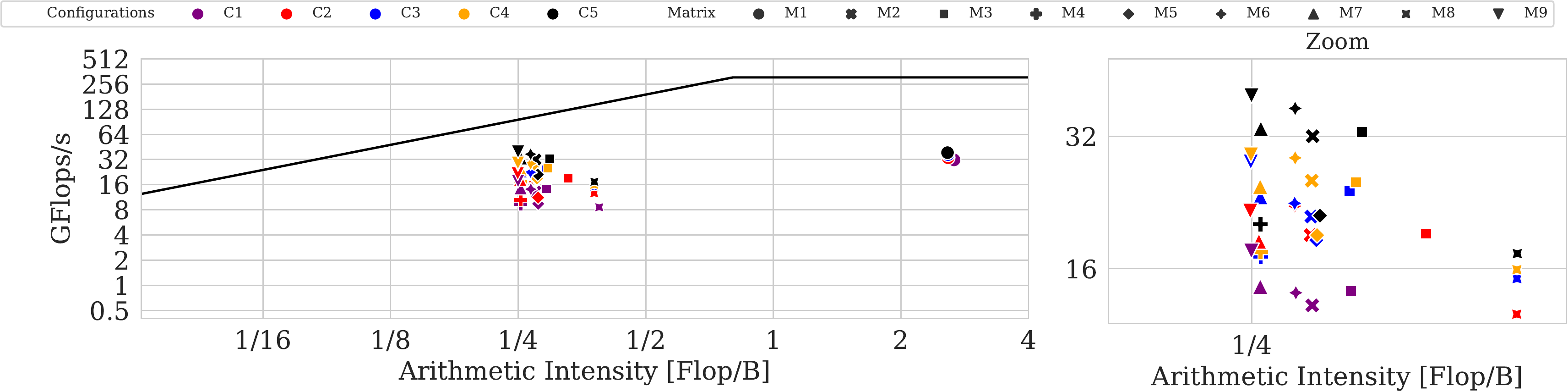}
    \caption{Roofline of our five configurations: our optimized configurations consistently yield an increase in throughput, up to $1.8\times$ on average.}
    \label{fig:sim_roofline}
\end{figure}

Secondly, enhancing the capabilities of having more memory requests in-flight (\textbf{C4}) improves performance by an average 79\% over the baseline. We further examine this in \Cref{fig:sim_roofline}, where we present the roofline for our configurations across all inputs. \spmv{} has low arithmetic intensity, consequence of relying on multiple memory requests to compute a single output value (\Cref{alg:spmv}). Hence, increasing the \glspl{mshr}' size empowers the \gls{cpu} to handle more misses without stalling, and increasing the \gls{rob} size allows for more independent load instructions to be queued. While these configurations are able to achieve only up $30\%$ of peak bandwidth utilization, increasing the capability of having more memory requests in-flight essentially doubles the bandwidth utilized w.r.t. \textbf{C1}.

Thirdly, having an interface optimized for bandwidth and one for latency in configuration \textbf{C5} improves performance by an additional 40\% over configuration \textbf{C4} and $2.17\times$ over the baseline. This is due to (i) better exploit the increased number of channels of \gls{hbm}2e, which is able to meet the high bandwidth requirements of streaming the sparse matrix without having to handle random accesses, (ii) the Open Row policy of \gls{hbm}2e which prevents \gls{dram} row buffers to be closed upon access, thus capitalizing on the fact that row buffer misses occur sporadically in streaming accesses, (iii) the Closed Row policy of \gls{ddr}5, which immediately closes the row buffer upon access and thus capitalizes on the bank precharge times not being on the critical path.

Finally, we observe that the size and locality of the input play a major role in how the proposed optimized architectures improve performance. First, if all the computational elements are able to fit in cache (\textbf{M1}), our optimizations yield limited benefits, which is a given of the fact that we mainly focus on optimizing for memory latency. Second, we observe that increasing the \gls{rob} size yields noticeable improvements in matrices with low Temporal Locality (\textbf{M2}, \textbf{M6}) while providing marginal speedups in higher locality inputs. For those inputs, increasing the \glspl{mshr}' size already yields a sufficient increase of in-flight memory requests, which are able to be filled before the \glspl{mshr}' are exhausted. 
\begin{figure}[t]
    \centering
    \includegraphics[width=\textwidth]{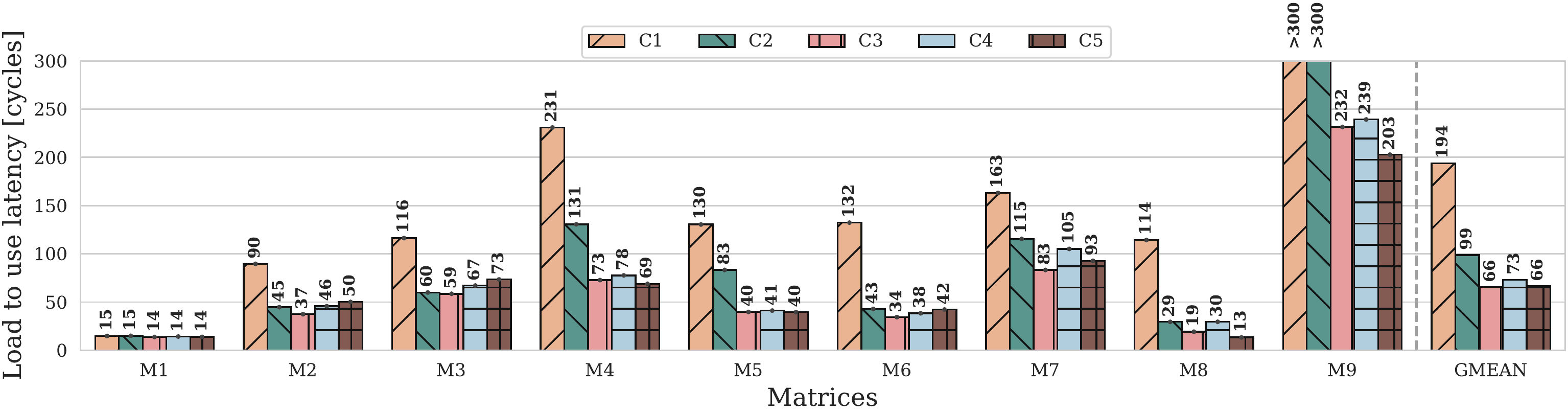}
    \caption{Load to use latency for our five configurations.}
    \label{fig:sim_ltul}
\end{figure}

\Cref{fig:sim_ltul} shows the average load-to-use-latency for each architecture and input matrix.
We observe that configuration \textbf{C5} yields an average reduction of $\approx$ $65\%$ in load-to-use latency. Moreover, we observe that such reduction is correlated with the matrix density, impacting more favorably sparser matrices (\textbf{M6}, \textbf{M7}, \textbf{M8}, \textbf{M9}). While all of our optimizations are effective in this scenario, adding private caches' prefetchers and increasing the \glspl{mshr}' size are the optimizations that yield the most improvement, that is, $49\%$ improvement using \textbf{C2} and an additional $33\%$ improvement using \textbf{C3}.

Overall, as outlined in \Cref{sec:optimizing}, increasing the number of in-flight memory requests and reducing the pressure on the memory controllers are effective optimizations for \spmv{}. Furthermore, our analysis shows that the most relevant improvements are obtained by optimizing memory interfaces for the type of access they ought to serve. In this setting,  we are able to best exploit the bandwidth of \gls{hbm}2e by performing streaming accesses exclusively on this interface and relinquishing all random accesses on an interface optimized for latency (\gls{ddr}5).

\section{Related  Work}\label{sec:soa}
To our knowledge, this is the first work that employs tree-based models to identify the most relevant hardware and input characteristics, starting from hardware and input-related metrics gathered from \glspl{pmc} and  matrices. There have been numerous characterization studies of Sparse Kernels in the literature, we now briefly discuss related prior work.

 There is a large body of research on comparisons across existing machines and instruction set architectures. Mantovani et al.~\cite{mantovani2020performance} evaluate thread and node scalability on the Marenostum 4 Supercomputer and Thunder X2 equipped systems. They compare performance and energy consumption, discuss microarchitectural choices (e.g., size of vector units or cache hierarchy) and conclude that \arm{} architectures are mature enough to debut in next-generation \gls{hpc} systems. 
 \fra{Armejach et al. \cite{hpc-tradeoffs-jos} evaluate various heterogeneous \gls{hpc} architectures employing different combinations of in-order and out-of-order cores as well as LLC sizes per core and conclude that scaling \glspl{llc} yields diminishing returns performance-wise, while having a noticeable impact on area and power utilization.}
 Poenaru et al.~\cite{poenaru2021evaluation} compare ARM (AWS Graviton2, Cavium TX2, Ampere Altra) and X86\_64 (Intel Cascade Lake, AMD Rome) platforms using both mini-applications and full-scale codes with different compilers (\cite{calore2020thunderx2,soria2021use} for TX2). Oliveira et al.~\cite{oliveira2021damov} evaluate ARM, PowerPC and X86 platforms for HPC via different metrics (e.g., \gls{gflops} per Watt, \gls{gflops} versus memory bandwidth or bytes per FLOP). For the specific case of ARM, they find that theoretical peak performances are much further away from sustained performances with respect to other mainstream platforms. 

Others have focused on using specific metrics for characterization. Bean et al.~\cite{bean2015g} characterize data movement of memory-intensive benchmarks by profiling an Intel Xeon platform with VTune. Then they run those benchmarks on a simulator to explore the impact of microarchitectural choices on the benchmark. Later, they classify the applications by the amount of spatial and temporal locality, and gather insights using \gls{mpki}, last to first level cache ratio, and operational intensity. Finally, they divide the problems into 6 classes, whose classification is dependent on how high/low the 3 metrics are. Cabezas et al.~\cite{cabezas2014extending} extend the roofline model with throughput, latency and cache capacity. They use instrumentation of the code to build a dependency graph, from which they estimate bottlenecks.  

Some works looked at the impact of sparse matrix formats. For example, Asgari et al.~\cite{asgari2021copernicus} discuss the performance implications of different sparse matrix formats. They focus on the decompression stage for non-standard formats using 20 matrices and provide a connection from matrix formats to applications. Alappat et al.~\cite{alappat2022execution} evaluate \spmv{} and Lattice quantum chromodynamics domain wall kernels on the A64FX platform comparing different matrix formats (CSR and SELL-C) to show which one is the most suitable. This work provides comprehensive architectural details and performance optimizations for A64FX. Finally, they build a model for single core execution time by taking into account architectural details like memory load time and instruction overlap.

We also find several Sparse kernel acceleration efforts in the literature. Some examples include \gls{fpga}-based~\cite{fowers2014high, grigoras2015accelerating,lin2013design,umuroglu2014energy,siracusa2020roofline,nguyen2022fpga}, dedicated hardware accelerators~\cite{nurvitadhi2015sparse,asgari2020alrescha,hegde2019extensor,hwang2020centaur,mishra2017fine, nurvitadhi2016hardware,pal2018outerspace,parashar2017scnn,qin2020sigma,zhang2021gamma, zhang2016cambricon,zhang2020sparch}, real \gls{pim} platforms~\cite{giannoula2022sparsep} or exploitation of software/hardware co-design~\cite{kanellopoulos2019smash, zhou2018cambricon,sadi2019efficient}. Finally, Lee et al.~\cite{lee2010debunking} find that, with appropriate optimizations, CPUs and GPUs can share similar performance regimes.

Regarding the use of \gls{ml} to predict performance, Chen et al.~\cite{chen2020characterizing} evaluate the scalability of \spmv{} from 1 to 4 cores, considering multiple matrices (1000+) from different categories. They build an \gls{ml} model to predict scalability from \gls{pmc} counters and different matrix metrics. While their approach is similar to ours, they only target scalability prediction in the context of a single CPU architecture and algorithm. Therefore, their feature extraction method is limited in scope and may not apply to different architectures and algorithms.

Therefore, none of these works perform a detailed characterization for a broad number of sparse kernels, inputs, and \glspl{cpu}. SpChar aims to fill this gap and to give guidance to software and hardware designers in developing solutions optimized for sparse computation. 

\section{Conclusion}
In this work, we present SpChar, a workload characterization methodology for general sparse computation. SpChar is based on tree-based models to identify the most relevant hardware and input characteristics, starting from hardware and input-related metrics gathered from \glspl{pmc} and matrices. SpChar's contributions are threefold: (1) it enables the characterization of sparse computation from the perspective of inputs, algorithms and architectures, (2) it determines what are the most impactful features for future architectures to excel in this field by gathering architectural insights, and (3) it creates a new analysis method to establish a characterization loop that could enable hardware and software designers to map the impact of architectural features to algorithmic choices and inputs. Our evaluation, which considers more than 600 matrices and various sparse kernels, determines that the biggest limiting factors for high-performance sparse computation are (1) the latency of the memory system, (2) the pipeline flush overhead due to branch misprediction and (3) the poor reuse of cached elements.  We test the SpChar methodology applied to the common use case of optimizing the \spmv{} kernel obtaining considerable performance improvements, hence SpChar is successfully able to create a characterization loop that enables software and hardware architects to map architectural features to inputs and algorithms, and optimize accordingly.

\section{Acknowledgment}
This research was supported by: the Spanish Ministry of Science and Innovation (MCIN) through contracts PID2019-107255GB-C21, PID2019-107255GB-C21 (MCIN/AEI/10.13039/501100011033) and the European Processor Initiative (EPI) under grant agreement No. 101036168. M. Siracusa has been partially supported by an FI fellowship (2022FI\_B 00969). 
 Finally, A. Armejach is a Serra Hunter Fellow.

\bibliographystyle{elsarticle-num} 
\bibliography{references}
\end{document}